\documentclass[times,doublespace]{simauth}
\usepackage{moreverb}
\usepackage[linkcolor=blue,urlcolor=blue,citecolor=blue, colorlinks,linktocpage,bookmarks,breaklinks,pdfstartview=FitH]{hyperref}

\newcommand\BibTeX{{\rmfamily B\kern-.05em \textsc{i\kern-.025em b}\kern-.08em
T\kern-.1667em\lower.7ex\hbox{E}\kern-.125emX}}

\usepackage{cite}
\usepackage{graphicx}

\usepackage{subfig}
\usepackage{paralist}
\usepackage{longtable}
\usepackage{multirow}

\def\mb#1{\mbox{\boldmath $#1$}}
\def\bb {\mb s}
\def\by {\mb y}
\def\blambda {\mb \lambda}
\def\btheta {\mb \theta}

\def \obs {^{\mbox{\tiny obs}}}
\def \noi {_{-i}}
\def \wi {_{1:n}}

\def \cvpostnoM {_{\mbox{\scriptsize post(-i)}}}
\def \QcvpostnoM {P\cvpostnoM(\btheta, \bb\wi | \by\obs\noi)}
\def \cvpostM {_{\mbox{\scriptsize post(-i)}}}
\def \QcvpostM {P\cvpostM(\btheta, \bb\noi | \by\obs\noi)}
\def \postfnoM {_{\mbox{\scriptsize post}}}
\def \QpostfnoM {P\postfnoM (\btheta, \bb\wi | \by\obs\wi)}

\def \QpostfM{P_{\mbox{\scriptsize post}} (\btheta, \bb\noi | \by\obs\wi)}
\def \nIS {^{\mbox{\scriptsize nIS}}}
\def \iIS {^{\mbox{\scriptsize iIS}}}

\def \postch {^{\mbox{\scriptsize Post}}}
\def \ghost {^{\mbox{\scriptsize ghost}}}

\def\dpois{\mbox{dpois}}
\def\ghostsub{_{\mbox{\scriptsize ghost}}}

\def\mci{^{(t)}}

\begin{document}

\runninghead{L.~LI, C.X.~FENG, and S.~QIU}

\title{Estimating Cross-validatory Predictive P-values with Integrated Importance Sampling for Disease Mapping Models}

\author{Longhai Li\corrauth\footnotemark[3], Cindy X. Feng\footnotemark[2],  and Shi Qiu\footnotemark[3]}

\address{\footnotemark[3] Department of Mathematics and Statistics, University of Saskatchewan, 106 Wiggins Rd, Saskatoon, SK, S7N5E6, Canada. 
\\ \footnotemark[2] School of Public Health, University of Saskatchewan, 104 Clinic Place, Saskatoon, SK, S7N5E5 Canada.}

\corraddr{Email:\texttt{longhai@math.usask.ca}}

\begin{abstract}
An important statistical task in disease mapping problems is to identify divergent regions with unusually high or low risk of disease.  Leave-one-out cross-validatory (LOOCV) model assessment is the gold standard for estimating predictive p-values that can flag such divergent regions. However, actual LOOCV is time-consuming because one needs to rerun a Markov chain Monte Carlo analysis for each posterior distribution in which an observation is held out as a test case. This paper introduces a new method, called integrated importance sampling (iIS), for estimating LOOCV predictive p-values with only Markov chain samples drawn from the posterior based on a full data set.   The key step in iIS is that we integrate away the latent variables associated the test observation with respect to their conditional distribution \textit{without} reference to the actual observation. By following the general theory for importance sampling, the formula used by iIS can be proved to be equivalent to the LOOCV predictive p-value. We compare iIS and other three existing methods in the literature with two disease mapping datasets. Our empirical results show that the predictive p-values estimated with iIS are almost identical to the  predictive p-values estimated with actual LOOCV, and outperform those given by the existing three methods, namely, the posterior predictive checking, the ordinary importance sampling, and the ghosting method by Marshall and Spiegelhalter (2003). 
\end{abstract}


\maketitle

\section{Introduction} \label{sec:intro}
\setcounter{section}{1}
\setcounter{page}{1}

In disease mapping, especially for mapping rare diseases, the observed disease count may exhibit extra Poisson variation. Hence, the standardized mortality ratios (SMRs), a basic investigative tool for epidemiologists, may be highly variable. Consequently, in maps of SMRs the most variable values (arising typically from low population areas) tend to be highlighted, masking the true underlying pattern of disease risk. To address this overdispersion issue, the field of disease mapping has flourished in the last decade with a variety of estimation methods and spatial models for latent levels of the model hierarchy. In particular, there have been many developments related to Bayesian hierarchical modelling which allow the estimate of the risk in a particular area to borrow strength from neighbouring areas where the disease risks are similar in order to produce maps of ``smoothed'' estimates of disease rates \cite{Besag1991,Clayton1992,Lawson2000, Best2005}.

There is great interest in being able to compute a predictive p-value for each unit (e.g., region or institution). A predictive p-value is the probability that the actual observation of a unit is larger (or smaller) than what is replicated (predicted/expected)  based on a model that has considered a set of factors (covariates); these predictive p-values measure the lack of fit of the observations to the model. A predictive p-value can be transformed using the standard normal quantile function. This transformation results in a generalized definition of residuals, called  quantile residual  \cite{dunn1996randomized}, which encompasses the traditional Pearson's residuals in normal regression as a special case.   Predictive p-values can be used for model checking and model diagnostics. Data modellers use these predictive p-values for two purposes. One such purpose is to check whether a proposed model fits adequately a dataset by comparing the predictive p-values (or their transformation) to a reference distribution, for example the uniform distribution \cite{dunn1996randomized, stern2000posterior, marshall2007identifying}. The other purpose is to discover systematic discrepancies (e.g., non-linearity effects, over-dispersion, or zero-inflation, or the necessity of including additional covariates) in order to suggest directions for improving a model \cite{dunn1996randomized,gelman2013two}.  Predictive p-values can be used (potentially) for a few practical purposes.  In an epidemiological context, predictive p-values can be used to \emph{identify} divergent units \cite{stern2000posterior,Marshall03,marshall2007identifying} for disease surveillance authorities.  These identified divergent units (\textit{e.g.} health regions or hospitals) will then be inspected closely by an expensive procedure.  This inspection may discover unknown factors associated with the unusually high or low disease rate that was identified by the divergent unit.  The predictive p-values could also be used in health and research resource allocation and other policy making procedures. For example, the units with unusually high disease rates may be allocated more research resources for finding underpinning causes.  Another potential use of predictive p-values is for ranking health care facilities or hospitals. Predictive p-values represent the ``residuals'' of the observed counts of a certain adverse event from the predictions based on a set of external factors, for example,  location factors, or the health conditions of patients or residents. With these external factors excluded, the predictive p-values (the residuals) reflect more accurately the internal quality of the health care facilities or hospitals than the original disease rates. 

There have been a number of methods proposed to compute such predictive p-values. The easiest method is to compute the posterior predictive p-value which is defined as the average of the tail probability of an observation (as a function of model parameters) with respect to the posterior distribution of parameters.  A similar posterior checking idea is used by \cite{gelman1996posterior}, where they define a p-value by averaging the tail probability of a discrepancy quantity (which summarizes the discrepancy of all observations rather than a single one) with respect to the posterior of model parameters.  The problem with these posterior predictive p-values is that the actual observations are used twice --- they are used to estimate the predictive distribution and used again to test the predictive distribution.   This leads to so-called optimistic bias or conservatism, where the actual observations appear more predictable by the model.  The consequence of this bias is that posterior predictive  p-values are concentrated around 0.5 rather than uniformly distributed on the interval (0,1); see \cite{gelman2013two}.  An appropriate method should have different datasets for estimating a predictive distribution and for testing the predictive distribution; cross-validatory methods separate a dataset into two parts for these two different tasks. \cite{stern2000posterior} suggests that leave-one-out cross-validation (LOOCV) predictive p-values be used as an alternative to posterior predictive p-values.  However, the actual LOOCV is time-consuming because one needs to rerun Markov chains in order to sample from each posterior distribution in which an observation is held out as a test case. Stern and Cressie \cite{stern2000posterior} suggest using the method of importance sampling (IS)~\cite{gelfand92bs} in order to compute the LOOCV p-values with MCMC samples from the posterior based on the full dataset. However, it is notorious that IS may have a large bias and variance \cite{peruggia1997variability, epifani2008case, vehtari2002bayesian}. Latent variables (or random effects) are often used in today's Bayesian models. Here, latent variables are random quantities that are associated with a subset of observations, for example a single observation. In contrast, we use ``parameters'' to refer to random quantities controlling the distribution of all observations.  For models with latent variables, a recent proposal for approximating LOOCV predictive p-values is the ghosting method \cite{Marshall03,marshall2007identifying}. The ghosting method discards the values of the latent variable associated with the test region in MCMC samples based on the full dataset and re-generates them from the distribution without reference to the actual observation of the test region. The ghosting method breaks the binding of the latent variable to the actual observation, reducing the optimistic bias; however, it does not correct for the optimistic bias in the model parameters.  Therefore, ghosting p-values cannot be proved in theory to be equivalent to the LOOCV p-values. 

Li et al.~\cite{li2015approximating} proposes a generic method called integrated importance sampling (iIS) for estimating LOOCV predictive quantities in latent variable models. iIS can be applied to estimate LOOCV predictive p-values. In this particular context,  iIS is closely related to the ghosting method, as iIS also discards and re-generates the values of the latent variable associated with the test region in each MCMC sample. However, iIS also considers the adjustment of the bias in the model parameters. Technically, iIS integrates the p-value and the likelihood of the observation of the test region with respect to the distribution of the latent variable without reference to the actual observation. Most importantly, the predictive p-values computed with iIS can be proven to be equivalent to the LOOCV predictive p-values by following the general theory for IS. ~\cite{li2015approximating} focuses on introducing the generic iIS formulae using rigorous and elaborate mathematical arguments, and demonstrates the method primarily by comparing LOOCV information criterion with other methods for computing information criterion, such as the DIC and WAIC.  However, precisely how to apply iIS to estimate LOOCV predictive p-values in disease mapping or similar models has not been presented clearly in \cite{li2015approximating}. The primary purpose of this paper is to provide a concrete description of the procedure of  applying iIS for computing predictive p-values for statisticians and analysts in relevant applied areas, as well as to compare the performance of iIS with existing methods for computing predictive p-values as reviewed above.  

This paper will be organized as follows. Section~\ref{sec:bmlv} reviews a Bayesian hierarchical model for disease mapping data. Section~\ref{sec:ppv} presents the details of how to estimate predictive p-values using the actual LOOCV, as well as four methods for computing predictive p-values with only MCMC samples from the posterior based on the full dataset; these methods are the posterior predictive checking method, the ordinary importance sampling method, the ghosting method, and the proposed iIS method. In Section~\ref{sec:examples}, we empirically investigate the four methods by comparing their predictive p-values to the actual LOOCV predictive p-values in two cancer count datasets collected in Scotland and Germany.  Our empirical results show that the LOOCV predictive p-values estimated with iIS are almost identical to those computed with actual LOOCV and are more accurate than those provided by the existing three methods. The article will be concluded in Section \ref{sec:conclusions} with a brief discussion of future work.

\section{A Bayesian Disease Mapping Model} \label{sec:bmlv}

We first consider a disease mapping dataset of Scotland lip cancer data, which was originally analyzed by \cite{clayton1987empirical} and was used by \cite{stern2000posterior}. The data represents male lip cancer counts (over the period of 1975-1980) in the $n = 56$ districts of Scotland. At each district $i$,  the data include these fields:
\begin{inparaenum}[(1)]
\item the number of observed cases of lip cancer, $y\obs_{i}$;
\item the number of expected cases, $E_{i}$,  calculated based on a standardization of  ``population at risk" across different age groups;
\item the standardized morbidity ratio ($S\!M\!R_i$) for the $i$th districts, $S\!M\!R_i \equiv y\obs_i/E_i$;
\item the percentage of the population employed in agriculture, fishing and forestry, $x_{i}$, used as a covariate; and
\item the group of IDs of districts neighbouring the $i$th district.
\end{inparaenum}
Table~\ref{tab:lpdata} shows the data for the first 6 districts. The full data for all 56 districts can be found from Table I of \cite{stern2000posterior}. \\

\begin{table}[htp]
\begin{center}
\vspace*{-10pt}
\caption{Scottish lip cancer data}\label{tab:lpdata}
\begin{tabular}{rlllllc}
  \hline
ID & District name & $y$ & $E$ &$S\!M\!R$ & $x$ & Neighbours \\
  \hline

  1 & Skye-Lochalsh &   9 & 1.38 & 6.52 &  16 & 5,9,11,19 \\
  2 & Banff-Buchan &  39 & 8.66 & 4.50 &  16 & 7,10 \\
  3 & Caithness &  11 & 3.04 & 3.62 &  10 & 6,12 \\
  4 & Berwickshire &   9 & 2.53 & 3.56 &  24 & 18,20,28 \\
  5 & Ross-Cromarty &  15 & 4.26 & 3.52 &  10 & 1,11,12,13,19 \\
  6 & Orkney &   8 & 2.40 & 3.33 &  24 & 3,8 \\
\hline
\end{tabular}
\end{center}
\end{table}

We consider here a typical Bayesian disease mapping model \cite{stern2000posterior} with a latent variable capturing the spatial correlation for a dataset (such as the Scottish lip cancer data). Let $\by=(y_1, \cdots, y_n)$ represent the vector of observed disease counts from $n$ geographical regions, where $\boldsymbol{E}=(E_1, \cdots, E_n)$ indicates the expected disease counts, and $\blambda=(\lambda_1, \cdots, \lambda_n)$ is a vector of relative risks (latent variables). Then, conditional on the expected counts and the relative risks, the response variables are assumed  independent and distributed as follows:
\begin{eqnarray}\label{eqn:model}
y_i|E_i,\lambda_i \sim \mbox{Poisson}(\lambda_iE_i).
\end{eqnarray}
To ensure $\lambda_i$ is positive, we model the logarithms of the relative risk, denoted by $\mb s_{1:n} =(s_1, \cdots, s_n)$, where $s_i=\mbox{log}(\lambda_i)$, as
\begin{eqnarray}\label{eqn:s}
\mb s_{1:n}  \sim N_{n}(\alpha + \mb X \boldsymbol{\beta},  \Phi\tau^{2}),
\end{eqnarray}
where $\mb X$ denotes the design matrix containing the values of covariate variables, $\boldsymbol{\beta}$ denotes the corresponding regression coefficients and $\Phi = (I_{n} - \phi C)^{-1}M$ is a matrix for capturing the spatial correlations amongst the $n$ districts, in which the elements of $C$ are: $c_{ij}= (E_{j}/E_{i})^{1/2}$ if areas $i$ and $j$ are neighbours, and $c_{ij} = 0$ if otherwise; the elements of $M$ are: $m_{ii}=E_{i}^{-1}$ and $m_{ij}=0$ if $i\not=j$; $\phi$ is a parameter measuring spatial dependence; $\Phi$ can be expressed as $M^{1/2}(I - \phi M^{-1/2}CM^{1/2})^{-1}M^{1/2}$. For positive definite $\Phi$, the range of $\phi$, $(\phi_{\min},\phi_{\max})$ is inverse of smallest and largest eigenvalues of $M^{-1/2}CM^{1/2}$ . The multivariate normal distributions with $\Phi$ as its covariance matrix are referred to as the \textit{proper conditional auto-regression (CAR) model}. Derived from the joint distribution in \eqref{eqn:s}, the conditional distribution of $\bb_{i}|\mb s_{-i}, \alpha, \beta, \phi$ is:
\begin{equation}
\bb_{i}|\mb s_{-i}, \btheta \sim N(\alpha + x_{i}\beta + \phi\sum_{j\in N_{i}}(c_{ij}(s_{j}-\alpha - x_{j}\beta)), \tau^{2} m_{ii}),
\label{eqn:pcarconds}
\end{equation}
where $N_{i}$ is the set of neighbours of district $i$, and $\mb s_{-i}$ denotes the collection of $\mb s_{j}$ except $\bb_{i}$: $\left\{s_j | j=1, \cdots, n, j\neq i \right\}$. We use $\btheta$ to collectively denote the model parameter vector $(\alpha, \beta, \tau, \phi)$.  For conducting Bayesian analysis, $\btheta$ is assigned independent and diffused priors:
\begin{eqnarray}
\alpha &\sim& N(0, 1000^{2}), \\
\beta &\sim& N(0,  1000^{2}), \\
\tau^{2} &\sim& \mbox{Inv-Gamma} (0.5, 0.0005), \\
\phi &\sim& \mbox{Unif}(\phi_{\min}, \phi_{\max}),
\end{eqnarray}
where $(\phi_{\min}, \phi_{\max})$ is the interval for $\phi$ such that $\Phi$ is positive-definite. When the number of regions is small, the Inverse-Gamma for $\tau^2$ may be better to replaced by a less restrictive prior such as the half-Cauchy \cite{gelman2006prior}. 

The above model is an example of a Bayesian model with unit-specific latent variables, which can be described symbolically as follows:
\begin{eqnarray}
y_{i} | \btheta, \bb_{i} &\sim& P_{y}(y_{i}|\mb \theta, \bb_{i}), \mbox{ for } i = 1,\ldots, n, \label{eqn:proby-gen} \\
\mb s_{1:n}|\btheta &\sim& P_{s}(\mb s | \btheta), \\
\btheta &\sim& \pi(\btheta) \label{eqn:probtheta-gen}.
\end{eqnarray}
Note that we omit the covariate variables (such as $E_{i}$ and $X_{i}$) for simplicity in the above generic model description. 

The above class of models includes many models that are widely used in different problems, including mixture models, factor analysis models, stochastic volatility models \cite{berg2004deviance, gander2007stochastic}, regression models with mixed effects \cite{gelman2006data}, and others.  We will demonstrate our new method (iIS) for predictive checks in the Bayesian disease mapping model. However, one should note that the method can be applied to all models that have the form as given by equations~\eqref{eqn:proby-gen} - \eqref{eqn:probtheta-gen}.

\section{Methods for Computing Predictive P-values} \label{sec:ppv}

\subsection{Posterior Predictive Checking} \label{subsec:pch}
Based on the models specified by equations~\eqref{eqn:proby-gen}-\eqref{eqn:probtheta-gen}, the \textbf{full data posterior} density of $(\bb_{1:n}, \btheta)$ given observations $\by_{1:n}\obs$ is given by:
\begin{equation}
\QpostfnoM=
\prod_{j=1}^n P_{y}(y_j\obs|s_j,\btheta) P_{s}(\bb_{1:n}|\btheta) \pi(\btheta)\,/\,C_1. \label{eqn:jointfull}
\end{equation}
where $C_1$ is the normalizing constant involving only $\by_{1:n}$.  In a posterior predictive assessment, one forms a posterior predictive density or mass function for replicated $y_{i}$ as follows:
\begin{eqnarray}
P\postfnoM(y_i|\by\obs\wi) &=& E\postfnoM\big[ P_{y}(y_i|\btheta, s_i)\big] \label{eqn:postpredy1}\\
&=&\int \int P_{y}(y_i|\btheta, s_i) \QpostfnoM ~ d\btheta d \bb\wi. \label{eqn:postpredy2}
\end{eqnarray}
In order to identify divergent observations for the model specified by~\eqref{eqn:proby-gen}-\eqref{eqn:probtheta-gen}, we apply the general posterior predictive checking method \cite{gelman1996posterior}  in order to look at the probability that the replicated $y_{i}$ is greater than observed $y_{i}\obs$ based on the posterior predictive distribution~\eqref{eqn:postpredy1}. Particularly, when $y_{i}$ is discrete,  the posterior predictive  p-value \cite{Marshall03, marshall2007identifying} is defined as follows:
\begin{equation}
\mbox{p-value}\postch( y_i\obs) = Pr\postfnoM (y_{i}>y_{i}\obs|\by\wi\obs) + 0.5Pr\postfnoM(y_{i}\obs|\by\obs\wi), \label{eqn:postpred-pv}
\end{equation}
where $Pr\postfnoM$ represents the probability of a set based on $P\postfnoM(y_{i}|\by\wi\obs)$.  Note that $y_{i}\obs$ is considered half in right tail for symmetry in two tails. For continuous $y_{i}$, the second term of \eqref{eqn:postpred-pv} is 0.   When this p-value is very close to 0 or 1, it indicates that the actual observed $y\obs_i$ falls on the tails of (ie, is unusual to) $P\postfnoM(y_{i}|\by\wi\obs)$, hence, there is a large discrepancy between the actual observation and the prediction (a distribution). The posterior predictive  p-value can be rewritten as an expectation of a function of $(\btheta, \bb_{i})$ with respect to \eqref{eqn:jointfull}:
\begin{eqnarray} \label{eqn:fullevalnoM}
\mbox{p-value}\postch( y_i\obs)&=& E\postfnoM (\mbox{p-value} (y_i\obs| \btheta,s_i)),
\end{eqnarray}
where  $\mbox{p-value} (y_i\obs| \btheta,s_i)$ is a p-value defined with respect to the predictive distribution of $y_{i}$ given parameters and latent variable:
\begin{eqnarray}\label{eqn:ppvpw}
\mbox{p-value} (y_i\obs | \btheta, s_i)  &=& Pr(y_i > y_i\obs | \btheta, s_i) + 0.5 Pr(y_i = y_i\obs | \btheta, s_i). 
\end{eqnarray}
Suppose we have obtained MCMC samples $\{(\btheta\mci, \bb\wi\mci); t = 1,\ldots, T\}$ from the full data posterior~\eqref{eqn:jointfull}. The posterior predictive  p-value~\eqref{eqn:fullevalnoM} for each observation $y_{i}\obs$ is computed as follows: 
\begin{eqnarray} \label{eqn:fullevalnoM-mc}
\widehat{\mbox{p-value}}\postch( y_i\obs)&=& \frac{\sum_{t}^{T} \mbox{p-value} (y_i\obs| \btheta\mci,s_i\mci ) }{T}.
\end{eqnarray}
For the poisson model given by equation~\eqref{eqn:model}, the p-value given parameters and latent variable is given by:
\begin{equation}
\mbox{p-value} (y_i\obs | \btheta, s_i) =\sum_{y>y_{i}\obs}\dpois(y|\lambda_{i}E_{i}) + 0.5\dpois(y_{i}\obs|\lambda_{i}E_{i}),
\end{equation}
where \dpois~is the Poisson probability mass function.

Posterior predictive checking uses the dataset twice: $y_{i}\obs$  is used to obtain the posterior predictive distribution~\eqref{eqn:postpredy1} of $y_{i}$, and is also used to test the goodness of~\eqref{eqn:postpredy1} which itself contains information from $y_{i}\obs$.  Using the dataset twice will introduce so-called optimistic bias in the predictive p-values, which means that the $y_{i}\obs$ appears better predictable by the model than it actually does. The consequence of this optimistic bias is that the posterior predictive  p-values are concentrated around 0.5 rather than uniformly distributed on the interval (0,1). This conservatism may not pose a serious problem if we only use the predictive p-values for discovering systematic discrepancies between a model and the dataset. However, when we use the predictive p-values to also check the goodness of fit of a model,   an inadequate model may appear to be a good fit due to conservatism; additionally, there is not a well-calibrated reference distribution to compare the predictive p-values against. When we use the posterior predictive p-values for identifying divergent regions, the observation in the tails of LOOCV predictive distribution (for which the observation itself is removed) may appear very compatible to the model.  An appropriate method should have different datasets for obtaining a predictive distribution and for testing the predictive distribution.  Cross-validatory methods separate a dataset into two parts for these two different purposes.

\subsection{Leave-one-out Cross-validatory Predictive P-value}\label{sec:cv}

Stern and Cressie \cite{stern2000posterior} proposed to use leave-one-out cross-validatory (LOOCV) method to obtain predictive p-values for identifying divergent regions in disease mapping. With the observation $y_{i}\obs$ left out as a test case, the cross-validatory posterior distribution $P\cvpostnoM(\btheta, \boldsymbol{s}_{1:n}|\by_{-i}\obs)$, is formed based on the observations except $y_{i}\obs$:
\begin{equation}
\label{eqn:MnoM}
P\cvpostnoM(\btheta, \bb_{1:n}|\by_{-i}\obs) = \prod_{j \not = i} P_{y}(y_j\obs|s_j,\btheta) P_{s}(\bb_{1:n}|\btheta) \pi(\btheta)\,/\,C_2,
\end{equation}
where $C_2$ is the normalizing constant involving only $\by_{-i}\obs$.  Note that we assume that the spatial relationships between $n$ locations are not lost, only that the value of $y_{i}\obs$ is omitted; that is, we treat the location as given information. The LOOCV predictive p-value for $y_{i}\obs$ is defined as the expectation of $\mbox{p-value} (y_i\obs| \btheta, s_i)$ \eqref{eqn:ppvpw} with respect to $P\cvpostnoM$: 
\begin{equation}
\mbox{p-value}(y_i\obs| \by\noi\obs)=E\cvpostnoM (\mbox{p-value} (y_i\obs| \btheta,s_i)). \label{eqn:ppv}
\end{equation}
Suppose we have obtained MCMC samples $\{(\btheta^{(t)}, \bb\wi^{(t)}); t = 1,\ldots, T\}$ from the LOOCV posterior~\eqref{eqn:MnoM}. The LOOCV predictive p-value \eqref{eqn:ppv} for each observation $y_{i}\obs$ is computed as follows:
\begin{eqnarray} \label{eqn:ppv-mc}
\widehat{\mbox{p-value}}^{\scriptsize \mbox{CV}}( y_i\obs)&=& \frac{\sum_{t}^{T} \mbox{p-value} (y_i\obs| \btheta\mci,s_i\mci ) }{T}.
\end{eqnarray}

When $y_{i}$ is continuous, Marshall and Spiegelhalter  \cite{marshall2007identifying} gives a proof that the LOOCV predictive p-value~\eqref{eqn:ppv} has a uniform(0,1) distribution when the distribution used to compute the p-value is indeed the true distribution generating $\by\wi\obs$.

The LOOCV predictive p-value can be rewritten in terms of the LOOCV predictive mass function of $y_{i}$:
\begin{equation}
\mbox{p-value}(y_i\obs| \by\noi\obs) = \sum_{y_{i}=y_{i}\obs+1}^{\infty}P\cvpostnoM (y_{i}|\by\noi\obs) + 0.5P\cvpostnoM(y_{i}\obs|\by\obs\noi), \label{eqn:cvpostpred-pv}
\end{equation}
where the LOOCV predictive mass or density function for $y_{i}$ is:
\begin{equation}
P\cvpostnoM(y_i|\by\obs\noi) = E\cvpostnoM\big[ P_{y}(y_i|\btheta, s_i)\big]. \label{eqn:cvpredpmf}
\end{equation}

Actual LOOCV requires $n$ Markov chain fittings (each may use multiple parallel chains), one for each observation. It is very time consuming, especially when the model is complex and $n$ is fairly large. Therefore, we are interested in estimating the expectation in~\eqref{eqn:ppv} for each testing observation $i=1,\ldots, n$ with samples of $(\btheta, \bb\wi)$ obtained with a single MCMC fitting based on the full data set; that is, with samples drawn from the full data posterior $\QpostfnoM$~\eqref{eqn:jointfull}.

\subsection{Non-integrated Importance Sampling} \label{subsec:is}

Gelfand et al. \cite{gelfand92bs} propose using importance sampling (IS) to estimate LOOCV prediction assessment quantities based on the full posterior, and Stern and Cressie \cite{stern2000posterior} propose using IS to estimate the LOOCV predictive p-value in disease mapping models as described in Section~\ref{sec:bmlv}. We will refer to this ordinary application of IS as non-integrated IS (nIS) to distinguish from the integrated IS that will be described in Section ~\ref{subsec:iIS}.   

For general and detailed discussions of importance sampling, one can refer to \cite{geweke1989bayesian, neal:1993, gelman1998simulating, liu:mcbook}; the following is a brief introduction.  Our goal is to find the expectation of a function $a(X)$ when $X$ has a probability density proportional to $f(x)$ (i.e., $f$ may be unnormalized); this expectation is denoted by $E_{f}\left(a(X)\right)$. If it is very expensive to draw samples from $f$,  we instead draw samples from an approximate distribution with a probability density proportional to $g(x)$. Let $W(x) = {f(x)}/{g(x)}$, called importance weighting function. Provided that the support of $g(x)$ is not smaller than that of $f(x)$, one can apply basic integration rules to show that the following identity holds:
\begin{eqnarray}
E_{f}\left(a(X)\right) & = & \dfrac{E_{g}\left(a(X)W(X)\right)}{E_{g} \left(W(X)\right)}. \label{eqn:is}
\end{eqnarray}
With \eqref{eqn:is}, one can use samples from $g$ to estimate the numerator and denominator and then obtain an estimate of $E_{f}(a(X))$. The intuition of the \textit{importance reweighting formula} \eqref{eqn:is} is that samples that are more compatible  with the target distribution $f$ (having larger ratio $f(x)/g(x)$) will be assigned more weight (and vice versa). 

Following \eqref{eqn:is}, we can estimate expectations with respect to $\QcvpostnoM$ in~\eqref{eqn:MnoM} by reweighting samples from $\QpostfnoM$~\eqref{eqn:jointfull} using the following identity:
\begin{equation}
 \label{eqn:isenoM}
 \mbox{p-value}(y_i\obs|\by\noi\obs) =
 \dfrac{E\postfnoM\big[ \mbox{p-value}(y_i\obs|\btheta, s_i) W_{i}\nIS(\btheta,\bb_{i})\big]}
 {E\postfnoM \big[W_{i}\nIS(\btheta,\bb_{i})\big]},
\end{equation}
where $W_{i}\nIS(\btheta,\bb_{i})$ is a value proportional to the ratio of  \eqref{eqn:MnoM} and \eqref{eqn:jointfull}: 
\begin{equation}
 \label{eqn:imwnoM} W_{i}\nIS(\btheta,\bb_{i}) = \frac {\QcvpostnoM} {\QpostfnoM} \times \frac{C_2}{C_1} = \frac{1}{P_{y}(y_i\obs|\btheta, s_i)}.
\end{equation}

We can estimate a LOOCV predictive p-value with Monte Carlo estimates of the numerator and denominator of \eqref{eqn:isenoM} with only MCMC samples from $\QpostfnoM$. If we have obtained MCMC samples $\{(\btheta^{(t)}, \bb\wi^{(t)}); t = 1,\ldots, T\}$ from the full data posterior \eqref{eqn:jointfull}, the nIS predictive p-value \eqref{eqn:isenoM} for each observation $y_{i}\obs$ is computed as follows:
\begin{equation}
 \label{eqn:isenoM-cv}
 \widehat{\mbox{p-value}}\nIS(y_i\obs) =
 \dfrac{\sum_{t}^{T}\Big[ \mbox{p-value}(y_i\obs|\btheta\mci, s_i\mci)\ W_{i}\nIS(\btheta\mci,\bb_{i}\mci)\Big]\Big/T}
 {\sum_{t}^{T} W_{i}\nIS(\btheta\mci,\bb_{i}\mci)\Big/T}.
\end{equation}

In theory, the IS estimate \eqref{eqn:isenoM} is valid and unbiased for almost all Bayesian models with latent variables. However, for models with latent variables (as well as many other models), the MCMC samples of  $\bb_i$ are largely bound to regions that fit the observation $y_i\obs$ well (since it is used to form the conditional distribution of $\bb_{i}$). Therefore, the distribution of $s_i$ marginalized from the full data posterior $\QpostfnoM$ may highly favour the region that fit the observation $y_i\obs$ well compared to the distribution of $s_i$ marginalized from the LOOCV posterior $\QcvpostnoM$, which is dissipated to a much larger region.  Importance reweighting \eqref{eqn:isenoM} attempts to reduce this optimistic bias in the full data posterior by assigning more weights to the samples less compatible with the observation $y_{i}\obs$.  However, this reweighting has the danger that the estimate \eqref{eqn:isenoM} is dominated by a single or a few very incompatible MCMC samples. This leads to the notorious instability problem of importance sampling; see \cite{peruggia1997variability, epifani2008case, vehtari2002cv,vehtari2015pareto}. 

\subsection{Ghosting Method} \label{subsec:ghost}

To break the binding of $\bb_{i}$ to the observation $y_{i}\obs$ in MCMC samples from the full data posterior, Marshall and Spiegelhalter \cite{marshall2007identifying} propose that at each MCMC sample $(\btheta, \bb\wi)$, the $\bb_{i}$ is discarded  and replaced with a re-generated $\bb_{i}$  from the distribution without reference to the actual observations $y_{i}\obs$, i.e., $P(\bb_{i}|\bb\noi,\btheta)$, as is the case when samples are drawn from the LOOCV posterior.  Probably because it is difficult to justify the role of such re-generated $\bb_{i}$ theoretically, they refer to the method as ``ghosting method''. Technically, the ghosting method estimates the LOOCV predictive p-value with the following equation:
\begin{equation}
\mbox{p-value}\ghost(y_i\obs) = E\ghostsub\left(\mbox{p-value}(y_i\obs | \btheta, s_i)\right), \label{eqn:ghostpv}
\end{equation}
where $\mbox{p-value}(y_i\obs|\btheta, s_i)$ is the same as in \eqref{eqn:ppvpw}, and the ``ghosting'' distribution of $(\btheta, \bb\wi)$ is defined as 
\begin{equation}
P_{\mbox{\scriptsize  ghost}}(\bb_{i}, \btheta) =P\postfnoM(\btheta, \bb\noi|\by\wi\obs)\times P(\bb_{i}|\bb\noi,\btheta),
\label{eqn:ghostdist}
\end{equation}
where $P\postfnoM(\btheta, \bb\noi|\by\wi\obs)$ is the marginalized distribution of $(\btheta, \bb\noi)$ given the full dataset~\eqref{eqn:postfullM}.   

Suppose we have obtained MCMC samples $\{(\btheta\mci, \bb\wi\mci); t = 1,\ldots, T\}$ from the full data posterior \eqref{eqn:jointfull}. To find a predictive p-value for each observation $y_{i}\obs$, the ghosting method will replace $\bb_{i}\mci$ \textit{temporarily} with a new $\tilde\bb_{i}\mci$ generated from $ P(\bb_{i}|\bb\noi\mci,\btheta\mci)$. With the new (``ghosting'') samples $\{(\btheta\mci, \tilde\bb_{i}\mci); t = 1,\ldots, T\}$, the ghosting predictive p-value is computed as follows:
\begin{eqnarray} \label{eqn:ghostpv-mc}
\widehat{\mbox{p-value}}\ghost ( y_i\obs)&=& \frac{\sum_{t}^{T} \mbox{p-value} (y_i\obs| \btheta\mci, \tilde \bb_i\mci ) }{T}.
\end{eqnarray}
Note that the original $s_{i}\mci$ should be retained for finding the predictive p-value for the other observations $y_{j}\obs$, $j\not=i$. 

The re-generation of $s_{i}$ makes the ``ghosting'' MCMC samples closer to samples from the LOOCV posterior, therefore, the ghosting predictive p-values are much closer to the LOOCV predictive p-values than the posterior predictive p-values, as we will see from our experimental results. However,  they are not equivalent in theory. This motivates us to find a new predictive p-value that is exactly equivalent to the LOOCV predictive p-value. 

\subsection{Integrated Importance Sampling} \label{subsec:iIS}

In this section, we propose to apply a new method called integrated importance sampling (iIS) \cite{li2015approximating} to estimate  the LOOCV predictive p-value.   iIS  also uses the idea that new $\bb_{i}$'s are re-generated from $P(\bb_{i}|\bb\noi, \btheta)$ to break the binding of $\bb_{i}$ to the actual observation $y_{i}\obs$. iIS employs the importance reweighting formula \eqref{eqn:is} to obtain a formula of predictive p-values that is exactly equivalent to the LOOCV predictive p-values \eqref{eqn:ppv} in theory.  The general formulae of iIS is presented in \cite{li2015approximating} with an elaborated derivation.  In what follows, we sketch the derivation of this iIS formula by focusing on the task of estimating the LOOCV predictive p-value \eqref{eqn:ppv}. We first rewrite the LOOCV predictive p-value \eqref{eqn:ppv} as an expectation of a function of $(\btheta, \bb\noi)$ by integrating $\bb_{i}$ away:
\begin{equation}
\mbox{p-value}(y_{i}\obs|\by\noi\obs) =\int \int  A (y_i\obs|\btheta, \bb\noi) P\cvpostnoM(\btheta, \bb\noi|\by\noi\obs) d\btheta d\bb\noi = E\cvpostnoM( A (y_i\obs|\btheta, \bb\noi))
 \label{eqn:cvevalM}
\end{equation}
where,
\begin{eqnarray}
A (y_i\obs| \btheta, \bb\noi) &=& \int \mbox{p-value} (y_i\obs| \btheta, s_i ) P(s_i |\bb\noi, \btheta)d s_i, \label{eqn:intA}\\
P\cvpostnoM(\btheta, \bb\noi|\by\noi\obs) &=&  \prod_{j \not = i} P_{y}(y_j\obs|s_j,\btheta) P_{s}(\bb\noi|\btheta) \pi(\btheta)\,/\,C_2, \label{eqn:postfM}
\end{eqnarray}
We will refer to the $A$ function as the \textbf{integrated p-value}. When using MCMC samples from the full data posterior $\QpostfnoM$,  we  discard $\bb_{i}$ in each sample.  The distribution of the retained parameters and latent variables $(\btheta,\bb\noi)$ can be derived by integrating $\bb_{i}$ out from the full data posterior \eqref{eqn:jointfull}, which results in the following expression:
\begin{equation}\label{eqn:postfullM}
 P\postfnoM(\btheta,\bb\noi|\by\obs\wi) =  \prod_{j\not=i} P_{y}(y_j\obs| s_j,\btheta) P(\bb\noi|\btheta) \pi(\btheta) \! \times \! P(y_i\obs|\btheta, \bb\noi) / C_{1},
\end{equation}
where the second factor is 
\begin{equation}
 \label{eqn:inpd}
 P(y_i\obs|\btheta, \bb\noi) = \int P_{y}(y_i\obs|s_i,\btheta)P(s_i|\bb\noi,\btheta)d\ s_i.
\end{equation}
We will refer to $P(y_i\obs|\btheta, \bb\noi)$ as the \textbf{integrated predictive density} of $y_{i}\obs$, since the $\bb_{i}$ in $P_{y}(y_{i}\obs|\bb_{i}, \btheta)$ is  integrated  out with respect to the distribution of $\bb_{i}$ given $\btheta$ without reference to the actual observation $y_{i}\obs$.  For the model specified in Section ~\ref{sec:bmlv}, $P(s_i|\bb\noi,\btheta)$ is the conditional normal distribution~\eqref{eqn:pcarconds}. Applying the importance reweighting formula \eqref{eqn:is} to estimate \eqref{eqn:cvevalM} with the sample of $(\btheta, \bb\noi)$ from $P\postfnoM(\btheta,\bb\noi|\by\obs\wi)$ \eqref{eqn:postfullM}, we obtain the iIS predictive p-value formula:
\begin{equation}
 \label{eqn:iseM} 
\mbox{p-value}(y_{i}\obs|\by\noi\obs) =
 \dfrac{E\postfnoM\big[ A(y_i\obs|\btheta,\bb\noi)\ W\iIS_{i}(\btheta,\bb\noi)\big]}
 {E\postfnoM \big[W\iIS_{i}(\btheta,\bb\noi)\big]},
\end{equation}
where the \textbf{integrated importance weight} $W_{i}\iIS$ is given by:
\begin{equation}
 \label{eqn:iswM} W_{i}\iIS (\btheta,\bb\noi) =
 \dfrac{\QcvpostM}{\QpostfM} \times \dfrac{C_{2}}{C_{1}} = \dfrac{1}{P(y_i\obs|\btheta, \bb\noi)}.
\end{equation}

The integration over $s_i$ in equations \eqref{eqn:intA} and \eqref{eqn:inpd} is the essential difference of iIS to IS. To apply iIS, it is therefore imperative to calculate the integral over $\bb_{i}$ in \eqref{eqn:intA} and \eqref{eqn:inpd}.  In some problems, they can be approximated with finite summation, or calculated analytically. Generally, we can use a Monte Carlo estimate by re-generating $s_i$ from $P(\bb_{i}|\bb\noi, \btheta)$.  

Suppose we have obtained the MCMC samples $\{(\btheta\mci, \bb\wi\mci); t = 1,\ldots, T\}$ from the full data posterior~\eqref{eqn:jointfull}.  The implementation procedure of finding the iIS predictive p-value for each observation $y_{i}\obs$ is described as follows. For each MCMC sample, we first generate two sets of  new $\bb_{i}$ from $P(\bb_{i}|\bb\noi\mci, \btheta\mci)$, denoted by $\{\tilde \bb_{i}^{(A,k)};k=1,\ldots, R\}$ and  $\{\tilde \bb_{i}^{(W,k)};k=1,\ldots, R\}$ respectively; these are used for estimating the integrated p-value $A(y_{i}\obs|\btheta\mci, \bb\noi\mci)$ and the integrated importance weight $W_{i}(y_{i}\obs|\btheta\mci, \bb\noi\mci)$ where: 
\begin{eqnarray}
\widehat{A}_{i}\mci &=& \dfrac{\sum_{k=1}^{R}\mbox{p-value} (y_i\obs| \btheta\mci, \tilde s_i^{(A,k)} )}{R}\\
\widehat{W}_{i}\mci &=& 1\Big/\frac{\sum_{k=1}^{R}P_{y} (y_i\obs| \btheta\mci,\tilde  s_i^{(W,k)} )}{R}.
\end{eqnarray}
The iIS predictive p-value \eqref{eqn:iseM} for the observation $y_{i}\obs$ is then computed as follows:
\begin{equation}
\widehat{\mbox{p-value}}\iIS(y_{i}\obs) = \dfrac{\sum_{t=1}^{T}\widehat{A}_{i}\mci\widehat{W}_{i}\mci \Big/T }{\sum_{t=1}^{T}\widehat{W}_{i}\mci\Big/T}. \label{eqn:iis-mc}
\end{equation}

From the above description, we can see that the ghosting method is a partial implementation of the iIS method.  In the ghosting method, only one new $\tilde s_{i}$ is generated from $P(s_{i}|s\noi\mci, \theta\mci)$ for each MCMC sample. Of course, this can be generalized to draw multiple $\tilde s_{i}$ and then find the mean of p-value estimated with these new $\tilde s_{i}$,  i.e., the $\widehat A_{i}\mci$. If we treat $\widehat W_{i}\mci$ to be an equal value for all MCMC samples in iIS method, the iIS estimate \eqref{eqn:iis-mc} is the same as this multiple-draw ghosting estimate \eqref{eqn:ghostpv-mc}.  The additional feature of iIS is to use $\widehat W_{i}\mci$ to \emph{reweight} each MCMC sample in order to correct the optimistic bias in $(\btheta\mci, \bb\noi\mci)$ due to the inclusion of the information of $y_{i}\obs$ in the full data posterior. After this correction, the quantity \eqref{eqn:iseM} (estimated by \eqref{eqn:iis-mc}) is exactly equal to the LOOCV predictive p-value \eqref{eqn:ppv} in theory. The amount of the optimistic bias in  $(\btheta\mci, \bb\noi\mci)$ depends on the flexibility, such as the number of parameters in $\btheta$. The amount may be small in simple models, but may be a more serious concern in more complex models. 

\section{Numerical Comparisons with Two Real Datasets}
\subsection{Lip Cancer Data in Scottland} \label{sec:examples}
In this section we will compare the four different methods for computing predictive p-values in the Scottish lip cancer data with respect to the Poisson model described in Section~\ref{sec:bmlv}. We used OpenBUGS through the R package \texttt{R2OpenBUGS} to run MCMC in order to obtain samples from  the full data posterior and the LOOCV posterior. For each \textit{MCMC fitting}, we ran two parallel chains, each with 15000 iterations; 5000 iterations were for burning in, and 10000 iterations were for sampling. 

We carried out the actual 56 cross-validatory MCMC fittings and used the MCMC samples of $(\btheta, s_i)$ to calculate the LOOCV predictive p-values  \eqref{eqn:ppv} for each of the 56 regions.  The LOOCV predictive p-values are shown in the column labeled as ``LOOCV'' of Table~\ref{tab:postinthree} in the appendix. We can see that some districts have very small and large LOOCV predictive p-values. For example, the LOOCV predictive p-values of district 2  and district 55 are 0.03 and 0.99, indicating that the lip cancer counts of these two districts are unusually higher and lower (respectively) than what is predicted by the assumed model described by~\eqref{eqn:proby-gen}-\eqref{eqn:probtheta-gen}, which considers two covariates ($x_{i}$ and $E_{i}$) and location effects.  Something unusual in these regions may have caused these high and low lip cancer counts.   For epidemiological practice, we can use a threshold such as 0.05 or 0.1 to determine whether a district is divergent or not. The divergent regions may receive further inspection by health authorities or research groups, which may lead to the discovery of additional factors (covariates) that have caused the unusually high or low disease rates.  The divergent districts with small p-values (high disease rate) may also be required to take actions to reduce the unusually high disease rates.   When the units are hospitals or health care facilities and $y_{i}$ represents the counts of a certain adverse event (such as mortality),  low predictive p-values (defined as upper tail in this article) indicate that the units may have unusually poor service quality because their adverse event incidence counts are much larger than what are expected/predicted based on a set of covariates including location. In this context, the hospitals or health care facilities may be required to improve their services in more than one aspect. 
\begin{figure}[htp]
\vspace{-10pt}
\centering
\subfloat[The LOOCV predictive PMF of $y_2$.]
{\includegraphics[width=0.4\textwidth, height = 0.25\textheight]{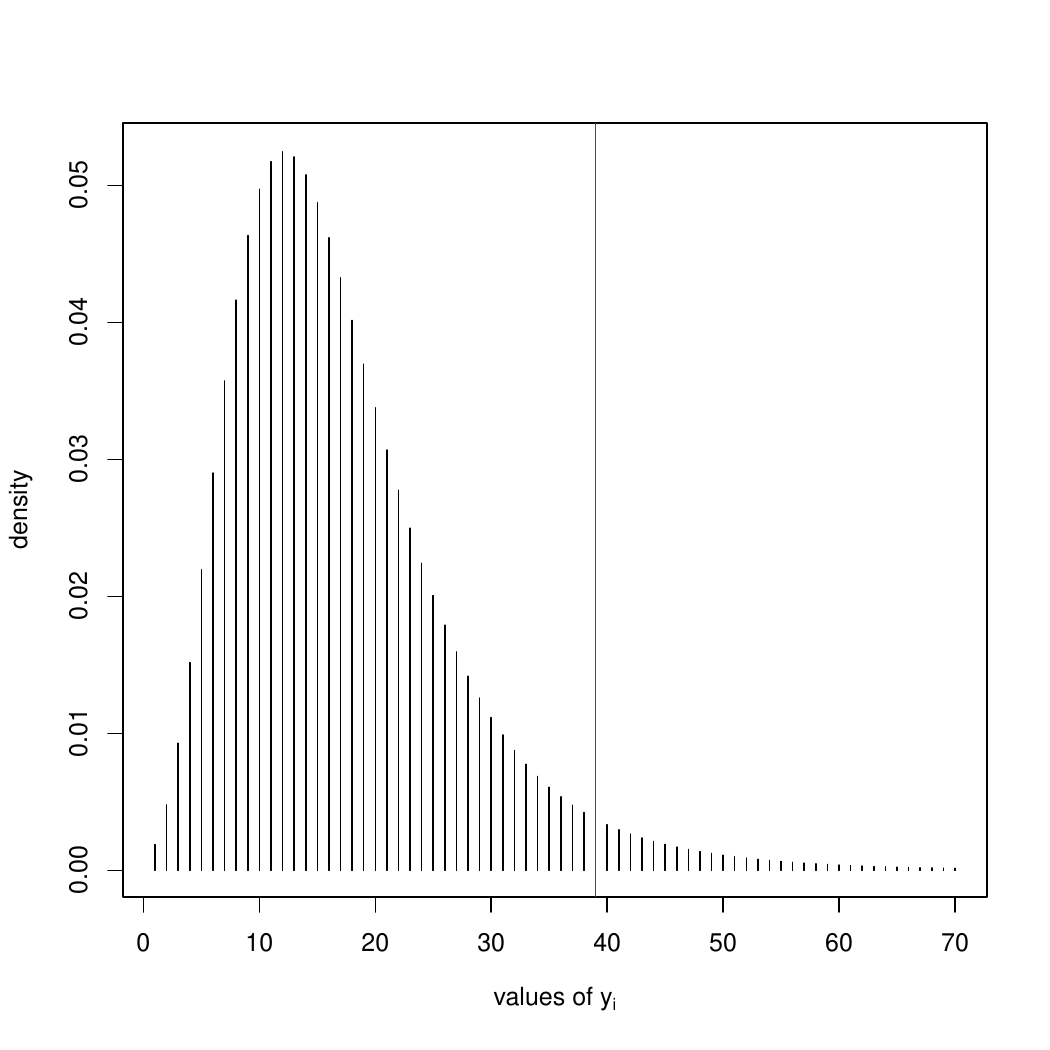}}~~
\subfloat[The full data posterior predictive PMF of $y_2$.]
{\includegraphics[width=0.4\textwidth, height = 0.25\textheight]{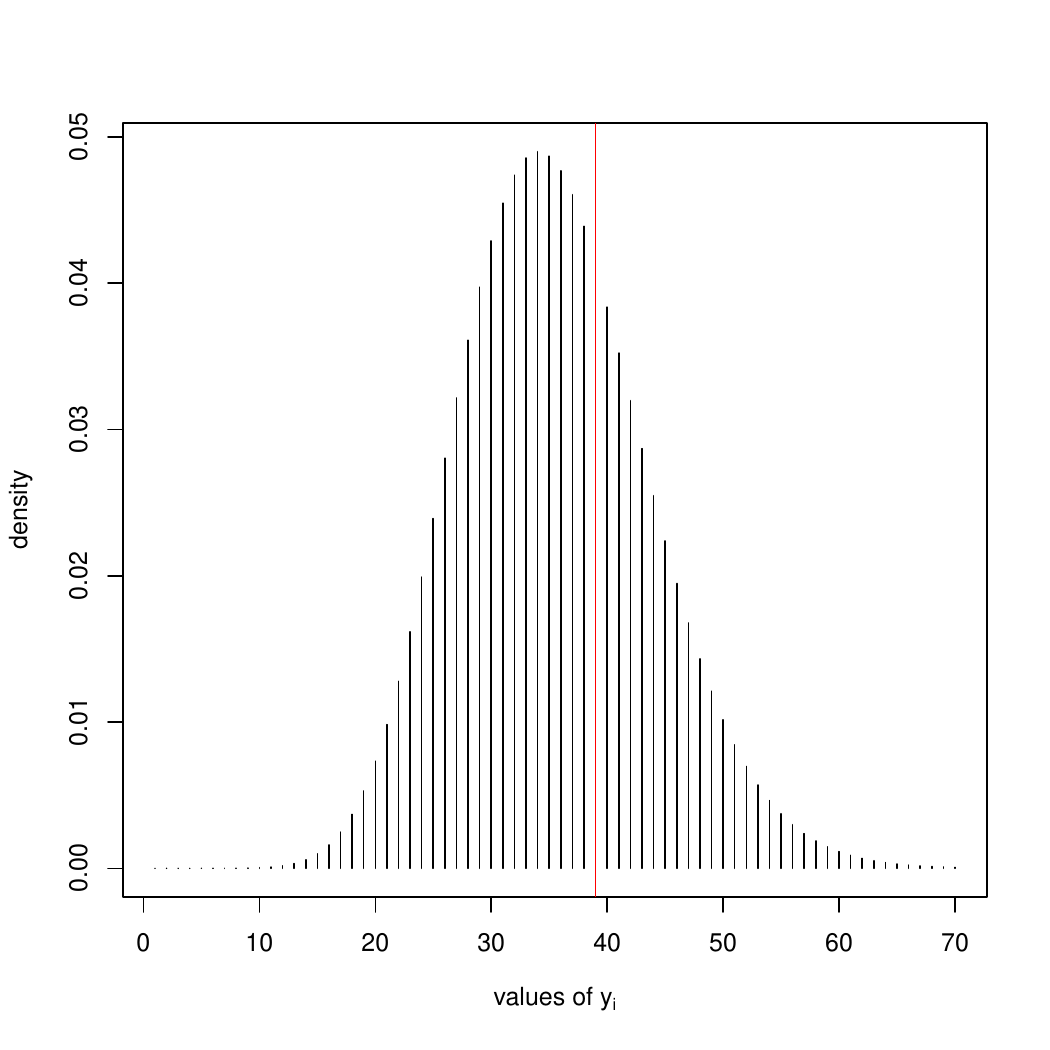}}

\caption{Comparisons of the predictive PMFs of $y_{2}$ (of district 2, Banff-Buchan) computed with the actual LOOCV~\eqref{eqn:cvpredpmf} and the posterior predictive checking method~\eqref{eqn:postpredy1}. The red vertical lines show the observed value $y\obs_2$.}
\label{fig:post_cv_y}
\end{figure}

We first explain the optimistic bias (conservatism) problem in posterior predictive  p-values.  Using the MCMC samples of $(\btheta, \bb\wi)$ from the full data posterior and the actual LOOCV posterior with $y_{2}\obs$ left out, we estimate the predictive  mass functions of the replicated $y_{2}$ with equations~\eqref{eqn:postpredy1} and~\eqref{eqn:cvpredpmf} respectively, for $y_{2}=0,\ldots, 70$. We compared the above two PMFs in Figure \ref{fig:post_cv_y} with red vertical lines indicating the actual observed values of $y\obs_2$ for district 2. We can see that although $y_{2}\obs$ lies on the tail of the LOOCV predictive distribution with a LOOCV p-value = 0.03 as seen from Table~\ref{tab:postinthree},  it is very plausible to the full data predictive PMF, which has a posterior predictive  p-value = 0.32. That is, the full data posterior, especially at $s_{2}$, is ``adapted'' to the actual observation $y_{2}$, whereas the LOOCV posterior is not since $y_{2}\obs$ is removed from the data. This bias arises because the full data posterior predictive distribution has indeed learned information of $y\obs_2$, hence it can predict $y_{2}\obs$ well. The consequence of optimistic bias is that the posterior predictive  p-values will concentrate more around 0.5 than the LOOCV predictive p-values.

\begin{figure}[htp]
\centering

\caption[Scatterplots of posterior p-values. ]{Scatterplots of predictive p-values against actual LOOCV predictive p-values. The districts in red are categorized differently by the full-data based predictive p-values and the LOOCV predictive p-values.}

\label{fig:pvlambdascatter}

\subfloat[Integrated IS (iIS)]
{\label{fig:pvlambdascatterd}\includegraphics[width=0.45\textwidth, height = 0.25\textheight, trim = 0 0 0 25pt]{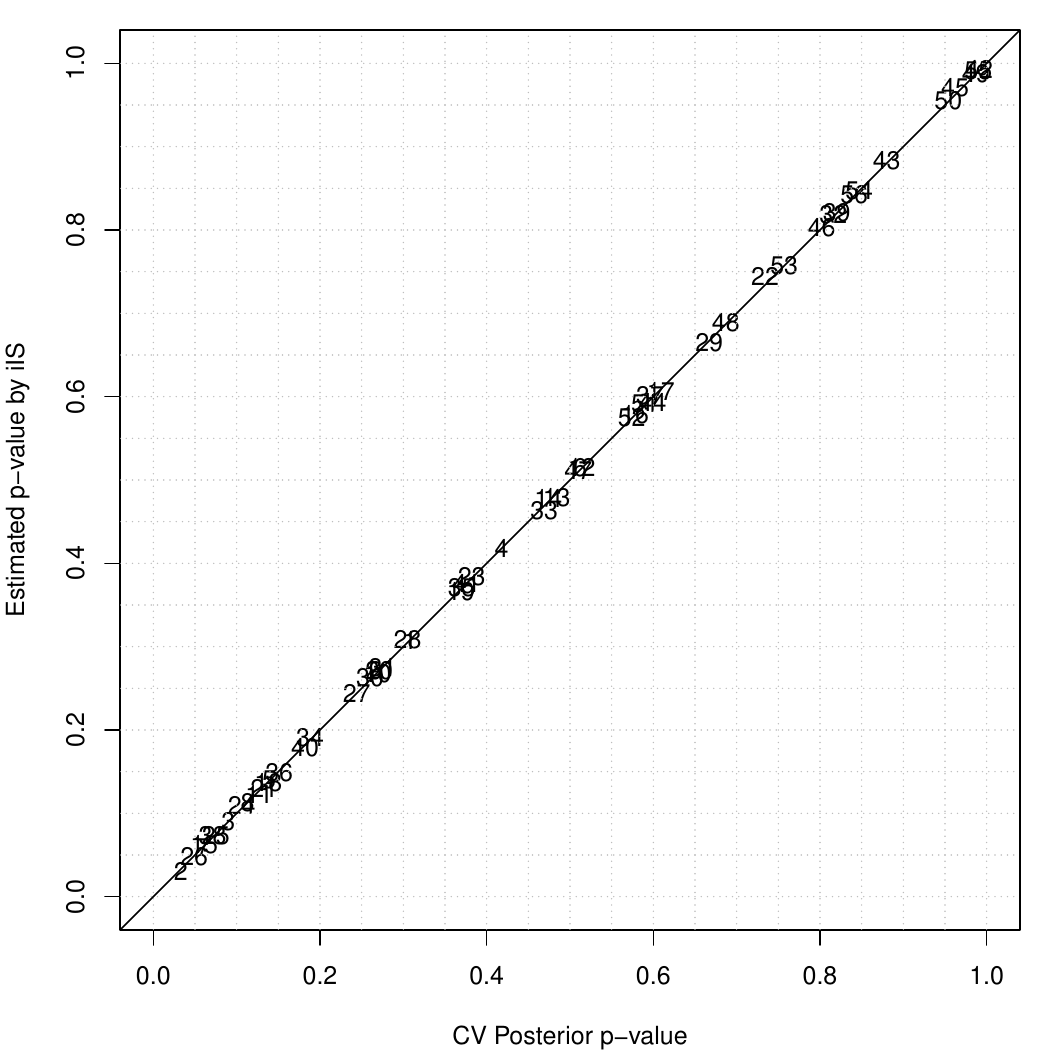} }
\subfloat[Posterior Predictive Checking]
{\label{fig:pvlambdascattera}\includegraphics[width=0.45\textwidth, height = 0.25\textheight, trim = 0 0 0 25pt]{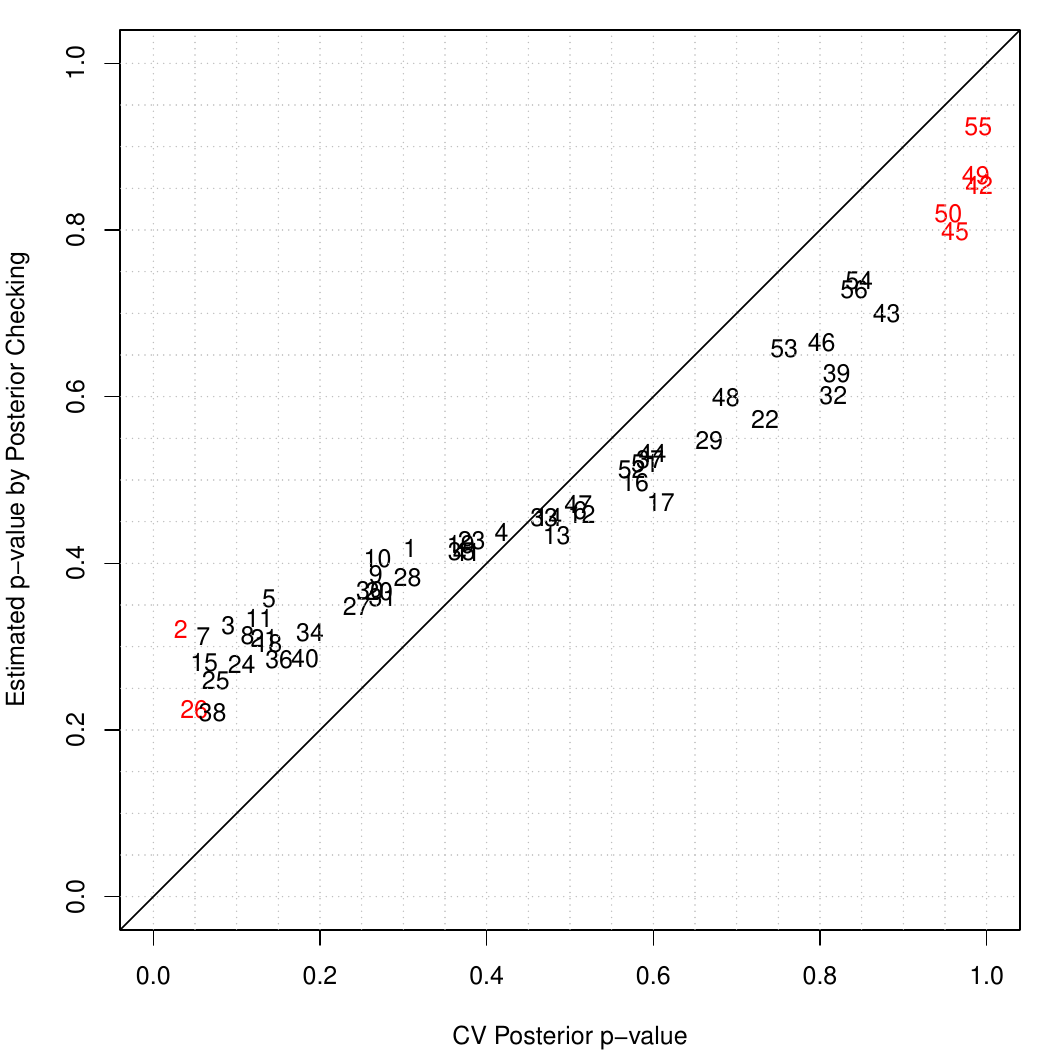}}

\subfloat[Ghosting method]
{\label{fig:pvlambdascatterb}\includegraphics[width=0.45\textwidth, height = 0.25\textheight, trim = 0 0 0 25pt]{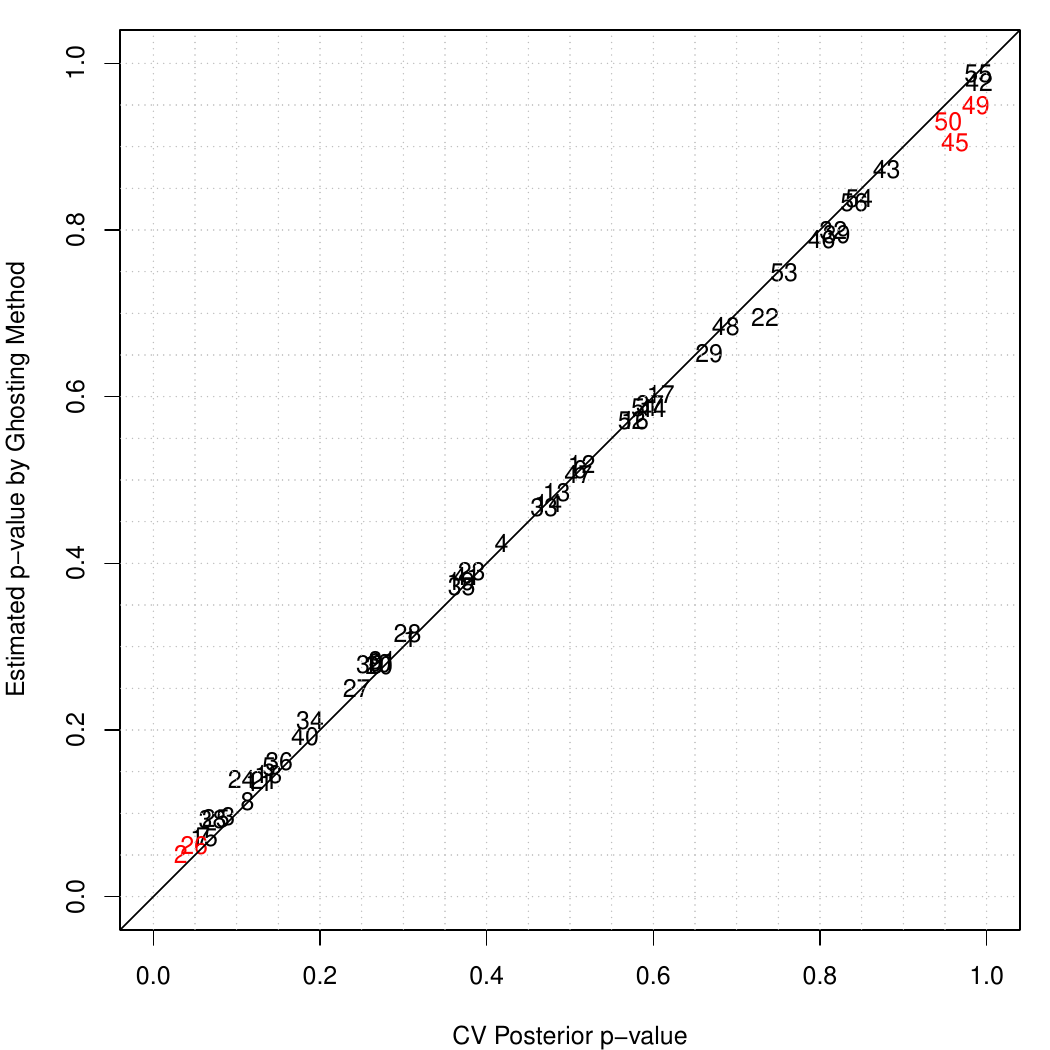}}
\subfloat[Non-integrated IS (nIS):I]
{\label{fig:pvlambdascatterc}\includegraphics[width=0.45\textwidth, height = 0.25\textheight, trim = 0 0 0 25pt]{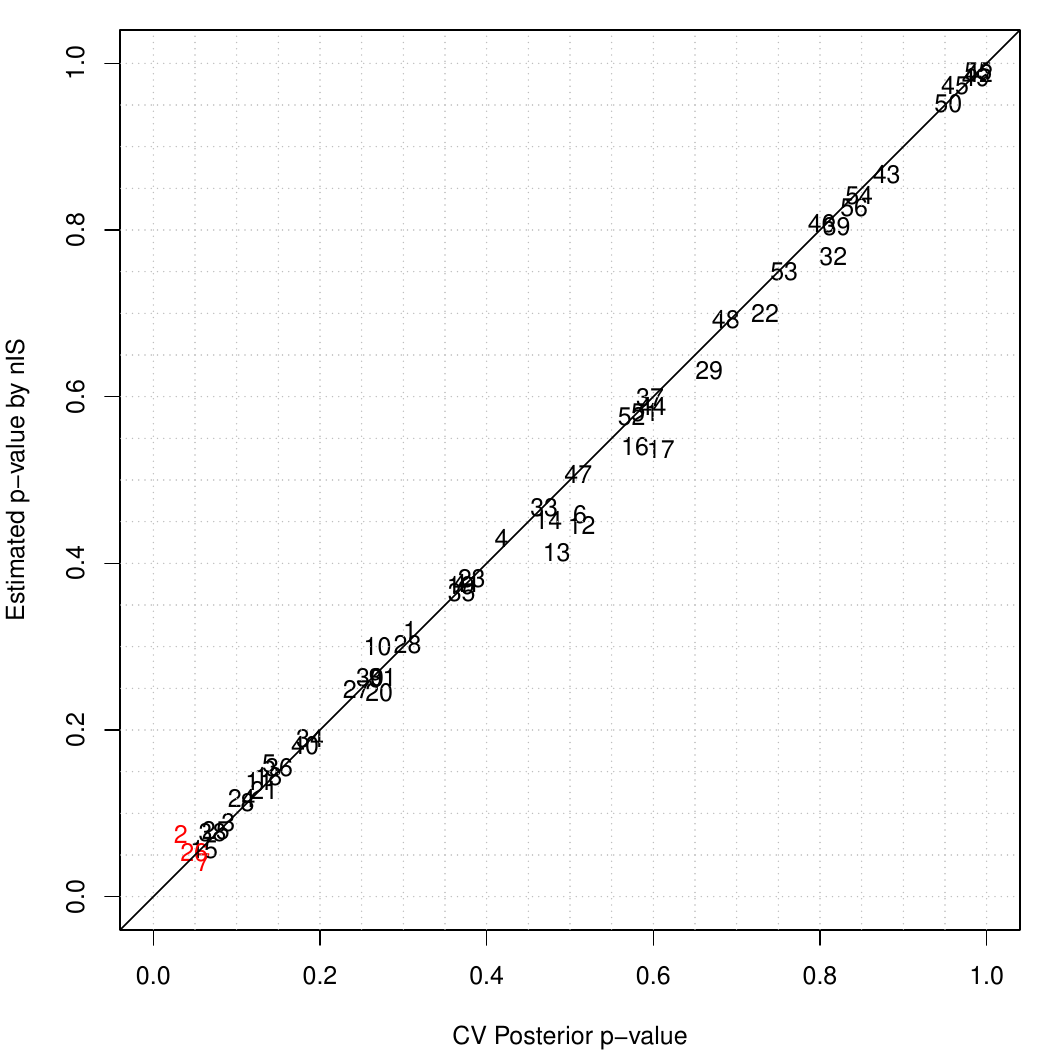}}

\subfloat[Non-integrated IS (nIS):II]
{\label{fig:pvlambdascattere}\includegraphics[width=0.45\textwidth, height = 0.25\textheight, trim = 0 0 0 25pt]{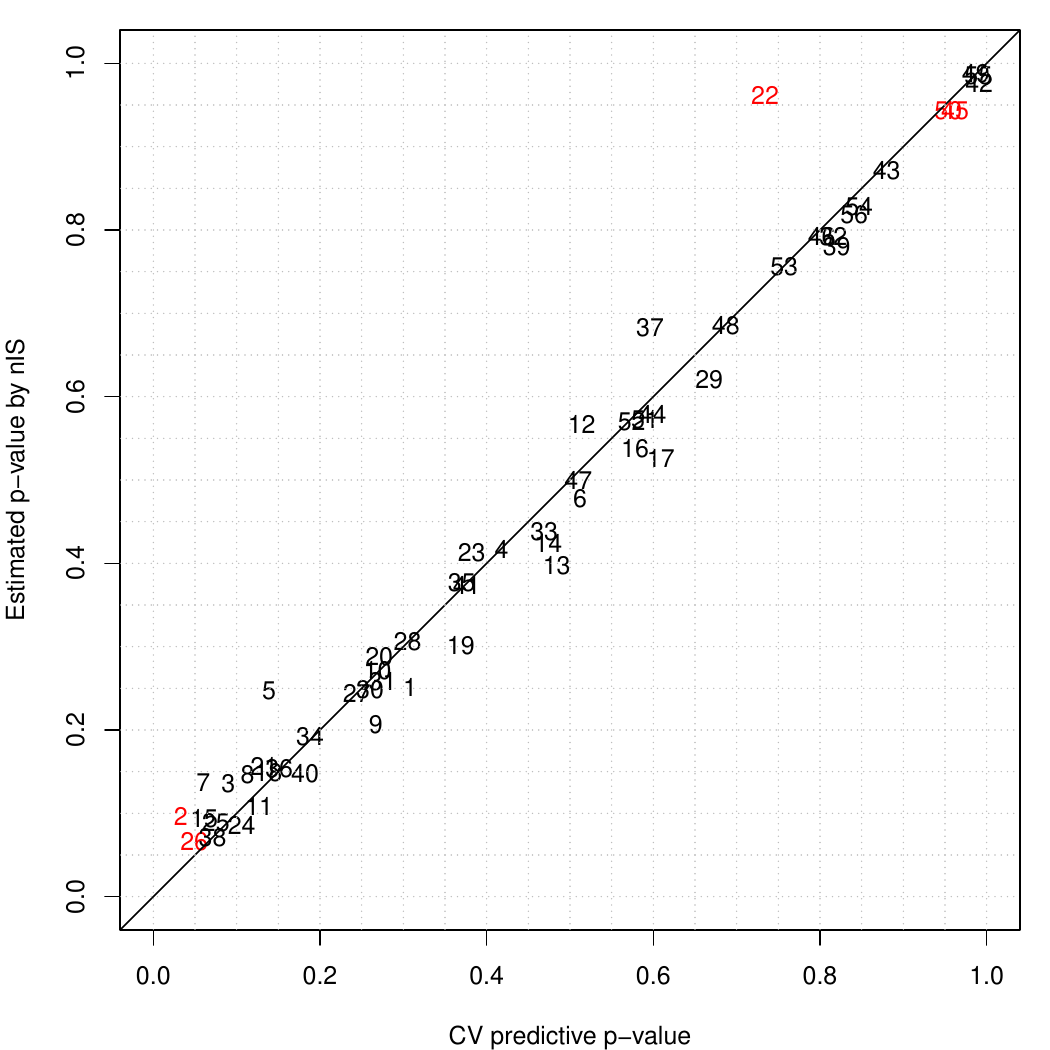}}
\subfloat[Non-integrated IS (nIS):III]
{\label{fig:pvlambdascatterf}\includegraphics[width=0.45\textwidth, height = 0.25\textheight, trim = 0 0 0 25pt]{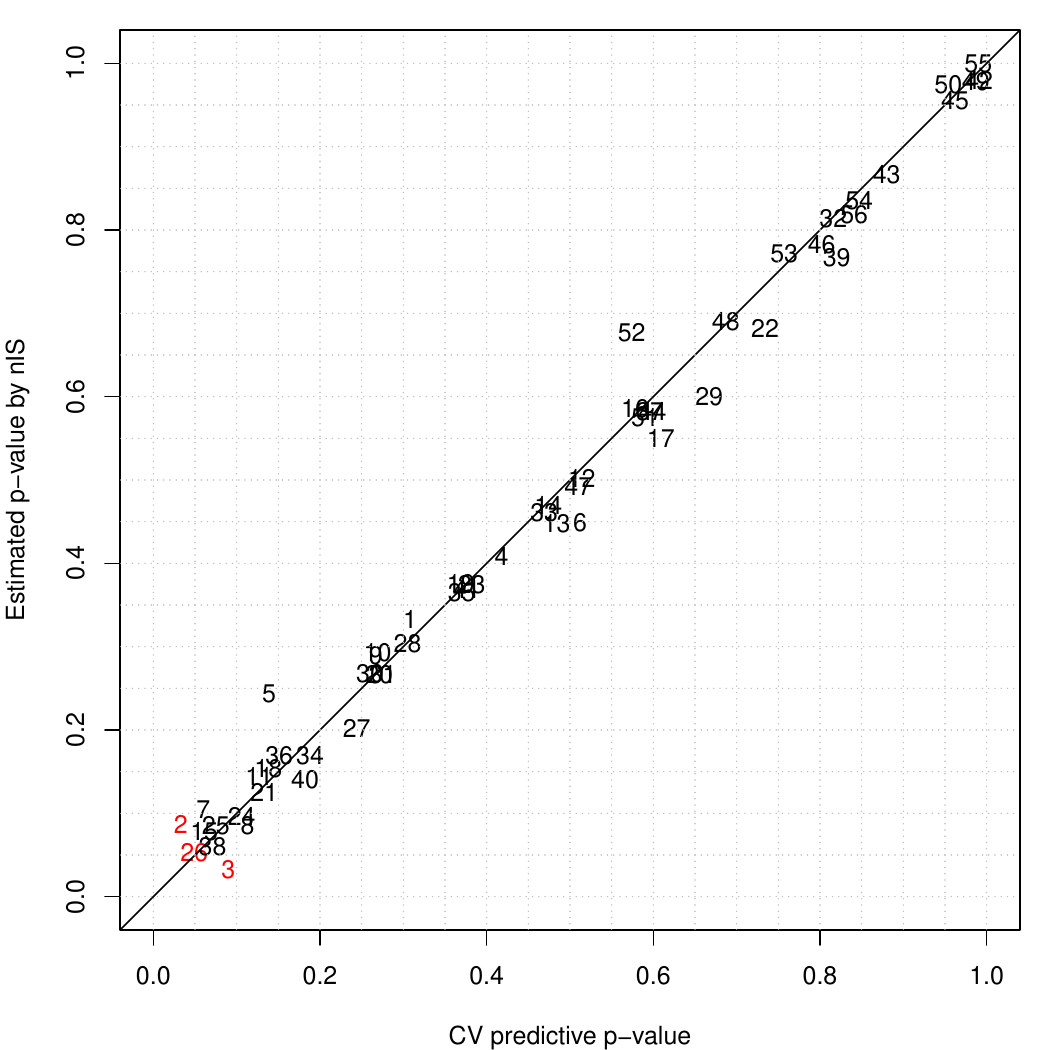}}

\end{figure}

We now compare the closeness of the p-values computed with  the posterior predictive checking, ghosting, nIS and iIS methods to the actual LOOCV predictive p-values. Figure~\ref{fig:pvlambdascatter} presents the scatterplots of  each set of predictive p-values against the actual LOOCV p-values. To demonstrate the instability of nIS, we show the scatterplots of three sets of nIS predictive p-values from three independent MCMC fittings for the same data and model. As depicted by Figure ~\ref{fig:pvlambdascatterd}, the iIS p-values are almost identical to the actual LOOCV p-values, with the scatterplot falling exactly on the diagonal line.  From Figure~\ref{fig:pvlambdascattera}, we see that the posterior predictive  p-values are more concentrated around 0.5 than the actual LOOCV predictive p-values, giving a scatterplot with an S-shape that is clearly distorted away from the diagonal line.  The ghosting method reduces the optimistic bias with regenerated $\bb_{i}$. However, there is still a slight S-shape in the scatterplot shown in  Figure~\ref{fig:pvlambdascatterb}. The case with the largest error is the district 45, for which the ghosting method gives a p-value 0.904, whereas the actual LOOCV p-value is 0.962. This is because that ghosting method does not correct for the optimistic bias in samples of model parameters due to using the data twice.  The scatterplots of the p-values generated by nIS (Figure ~\ref{fig:pvlambdascatterc}-\ref{fig:pvlambdascatterf}) do not show a visible S-shape. Nevertheless, the plot shows many jitters around the diagonal line, indicating high variability in the nIS p-values. In addition, we see that for the same data and model, three sets of nIS predictive p-values are highly varied for some regions.  The integrations with respect to the $P(\bb_{i}|\bb\noi, \btheta)$ in~\eqref{eqn:intA} and \eqref{eqn:iswM} help reduce this variability. 

The raw discrepancies between the full-data based predictive p-values and the LOOCV predictive p-values seem small as they appears in Figure~\ref{fig:pvlambdascatter}. However, these small discrepancies may lead to wrong decisions with serious implications when they are used to categorize the districts into different pools; whether for identifying divergent districts, or for other practical purposes.  Suppose we want to categorize the 56 districts into three pools by cutting the predictive p-values with 0.05 and 0.95.  In Table~\ref{tab:postinthree}, we embolden the predictive p-values that result in different categorization (mis-categorization) of the districts than when we cut the LOOCV predictive p-values. In  Figure~\ref{fig:pvlambdascatter}, we highlight the mis-categorized districts for each method in red.  We see that the posterior predictive checking and ghosting predictive p-values produce 7 and 5 mis-categorized districts respectively; the nIS predictive p-values produce 3 or 5 mis-categorized districts from three different MCMC fittings. The iIS predictive p-values also gives exactly the same categorization that is given by the LOOCV predictive p-values.  

To quantify the discrepancies (errors) between each set of full-data based p-values to the actual LOOCV p-values, we used a relative error quantity defined to be:
\begin{equation}\label{eqn:reer}
\mbox{relative error} = (1/n)\sum_{i=1}^{n}\frac{|\hat p_i - p_i|}{\min(p_i, 1-p_i)} \times 100,
\end{equation} 
where $\hat {p}_{i}$ is an estimate of the actual LOOCV p-value ${p}_{i}$. This measure puts more weight on the error between $\hat p_i$ and $p_i$ when $p_i$ is very small or very large, for which we demand more on the accuracy of an estimate than when $p_i$ is close to $0.5$ in the problem of identifying divergent units. A similar measure with only $p_{i}$ in the denominator was suggested in~\cite{marshall2007identifying}. Here, we modify the denominator to consider the errors associated with large p-values because large p-values also signify divergent units. Table \ref{tab:ppvtab} shows the averages of these relative errors over 100 independent MCMC fittings based on the same data and model. Clearly, iIS outperforms all the other competing methods. iIS not only has a smaller mean relative error, but also smaller variability than other methods.  The relative errors in the posterior predictive  p-values are clearly larger than all other methods.  The ghosting method is stable in the 100 replicated MCMC fittings, but has a relative error that is consistently larger than iIS, indicating that the S-shape in ghosting p-values appear in all of the 100 replicated MCMC fittings. nIS gives slightly better estimates of small or large (extreme) p-values better than the ghosting method; however, as shown by the three sets of nIS predictive p-values in Figure~\ref{fig:pvlambdascatter}, there is non-neglible variability in nIS predictive p-values, which is also reflected in the significantly larger standard deviation than the iIS and ghosting methods in relative errors in the 100 independent MCMC fittings.

\begin{table}[ht]
\centering
\caption{Comparison of relative errors of the estimated predictive p-values to the actual LOOCV predictive p-values. The numbers outside the brackets show means of relative errors in 100 independent MCMC fittings. The numbers in the brackets show the standard deviations of relative errors in the 100 MCMC fittings. Abbreviations:  PCH: posterior predictive checking, GHO: Ghosting, nIS: non-integrated importance sampling, iIS: integrated importance sampling.}\label{tab:ppvtab}
\begin{tabular}{llll}
  \hline
  iIS & nIS & GHO & PCH \\
  \hline
  1.501(0.210) & 12.481(\textbf{1.586}) & 19.212(0.359) & 160.580(1.101) \\
   \hline
\end{tabular}
\end{table}


To compare the computational efficiency of different methods for computing predictive p-values, we also recorded the execution time for the process of computing p-values. We considered time consumed in two parts: generating MCMC samples of $(\btheta, s_i)$, and computing predictive p-values from these samples. As shown in Table \ref{tab:time_y}, the time spent on MCMC fittings using the LOOCV method is about 56 times as large as the time used by the other methods; this is because LOOCV requires 56  MCMC fittings with each district removed, whereas the other methods need only one MCMC fitting given the full dataset.  iIS requires additional time for computing p-values compared to the other three full-data based methods; this is because of the required additional computations for finding the integrated p-value and the integrated importance weight.  The total computing time of iIS is roughly 8 times as large as that of nIS, and 1.6 times as large as that of the ghosting method.  Although iIS requires more time than these methods, the increased accuracy in the iIS predictive p-values may be necessary when the predictive p-values are used in practical decisions as we discuss above. \begin{table}[htp]
\centering

\caption{Comparison of computation time (in seconds). (Abbreviations: LOOCV: actual cross validation, PCH: posterior predictive checking, GHO: Ghosting, nIS: naive importance sampling and iIS: integrated importance sampling).}\label{tab:time_y}
\begin{tabular}{rrrrrr}
  \hline
  & LOOCV & PCH & nIS & GHO & iIS   \\
  \hline
  MCMC fitting & 1138 & 20 & 20 & 20 & 20   \\
 Computing p-values & 1& 1 & 1 & 84& 144   \\
  Total & 1139 & 21 & 21 & 104 & 164   \\
   \hline
\end{tabular}
\end{table}

\subsection{Larynx Cancer Data in Germany}

In this section, we compare the four methods for computing predictive p-values using MCMC samples from the full data posterior in a larger dataset of cancer mortality counts collected in $N= 544$ districts of Germany from the year of 1985 to 1990. The details of the dataset is given by \cite{becker1997atlas}.  From this dataset, the larynx cancer mortality counts  are denoted by $y_{i}$ and the expected counts, $E_{i}$, are calculated by accounting for the population and age distribution in each district. The level of smoking consumption, $x_{i}$, in each district (used as a covariate for $y_{i}$),  and the neighbouring information of the 544 districts were extracted and used by \cite{held2005towards} and \cite{rue2005gaussian}.  We downloaded a dataset containing the above variables including the neighbouring relationships from the accompanying website (\url{http://www.r-inla.org/examples/volume-1}) for the R package \texttt{INLA}.    We use the same model as for the lip cancer dataset (as described in Section~\ref{sec:bmlv}) for modelling this dataset with OpenBUGS.  Because the dataset is much larger than the previous example, we decided to run MCMC for 30000 iterations. MCMC convergence was diagnosed with usual tools such as Rhat as well as by visual inspection of MCMC traces. 

We ran a single MCMC fitting (with two independent chains) given the full dataset with 544 districts, and applied the four methods described in Sec. \ref{sec:ppv} to compute predictive p-values.  In applying iIS, we drew two additional sets of 50 samples of $\bb_{i}$ for estimating the integrated p-value and integrated predictive density for each MCMC sample, as well as for each observation $y_{i}$.  We ran the actual 544 cross-validatory MCMC fittings for the dataset with each observation $y_{i}\obs$ removed, and then calculated the LOOCV predictive p-values. This computation was very intensive since each MCMC fitting takes roughly 2.5hrs. We used a computer cluster to parallelize the 544 MCMC fittings in order to obtain the LOOCV p-values.  The comparison of these computation times is presented in Table~\ref{tab:time_y_larynx}. For LOOCV, we show the total computation time for running the 544  MCMC fittings and computing p-values.  From Table~\ref{tab:time_y_larynx}, we see that the additional time for iIS  to compute p-values compared to the nIS and the posterior predictive checking methods (from 2 seconds to $\approx$ 9 mins) becomes relatively small because the MCMC fitting itself requires much more time (2.5hrs per MCMC fitting, and a total of 1333hrs for all 544 MCMC fittings). Additionally,  we see that the extra time (about 4mins) for iIS for computing p-values compared to the ghosting method, which draw only one additional $\bb_{i}$ for each MCMC sample and each observation, is also very small. Therefore, the total times for the four methods are almost the same since MCMC fitting has dominated the total times, and all of the four methods gain a huge time saving compared to LOOCV with a relative ratio less than 1.73/1000 (iIS).   Finally, we point out that the p-value computation with iIS can also be paralleled with a computer cluster if one desires even faster computation. 

\begin{table}[ht]
\centering
\caption{Comparison of computation time in larynx cancer example. (Abbreviations: LOOCV: actual cross validation, PCH: posterior predictive checking, GHO: Ghosting, nIS: naive importance sampling and iIS: integrated importance sampling).}\label{tab:time_y_larynx}
\begin{tabular}{lrrrrr}
  \hline
 &LOOCV & PCH & nIS & GHO & iIS \\ 
  \hline
MCMC fitting (seconds) & 4.8$\times10^{6} $ & 7816 & 7816 & 7816 & 7816 \\ 
  Computing p-values (seconds)  & 2 & 2 & 2 & 284 & 522 \\ 
  Total (seconds) & 4.8$\times10^{6}$  & 7818 & 7818 & 8100 & 8338 \\ 
  Total (hours) & 1333 &  2.17 & 2.17 & 2.25 & 2.32 \\
  Total (relative to CV) & 1 & 0.162\% &0.162\% &0.168\% & 0.173\%\\
   \hline
\end{tabular}
\end{table}

\begin{figure}[htp]
\caption[Scatterplots of posterior p-values. ]{Scatterplots of predictive p-values against actual LOOCV predictive p-values in Larynx cancer example.} \label{fig:larynx-pvalues}
\centering
\includegraphics[width = 0.85\textwidth, height = 0.55\textheight]{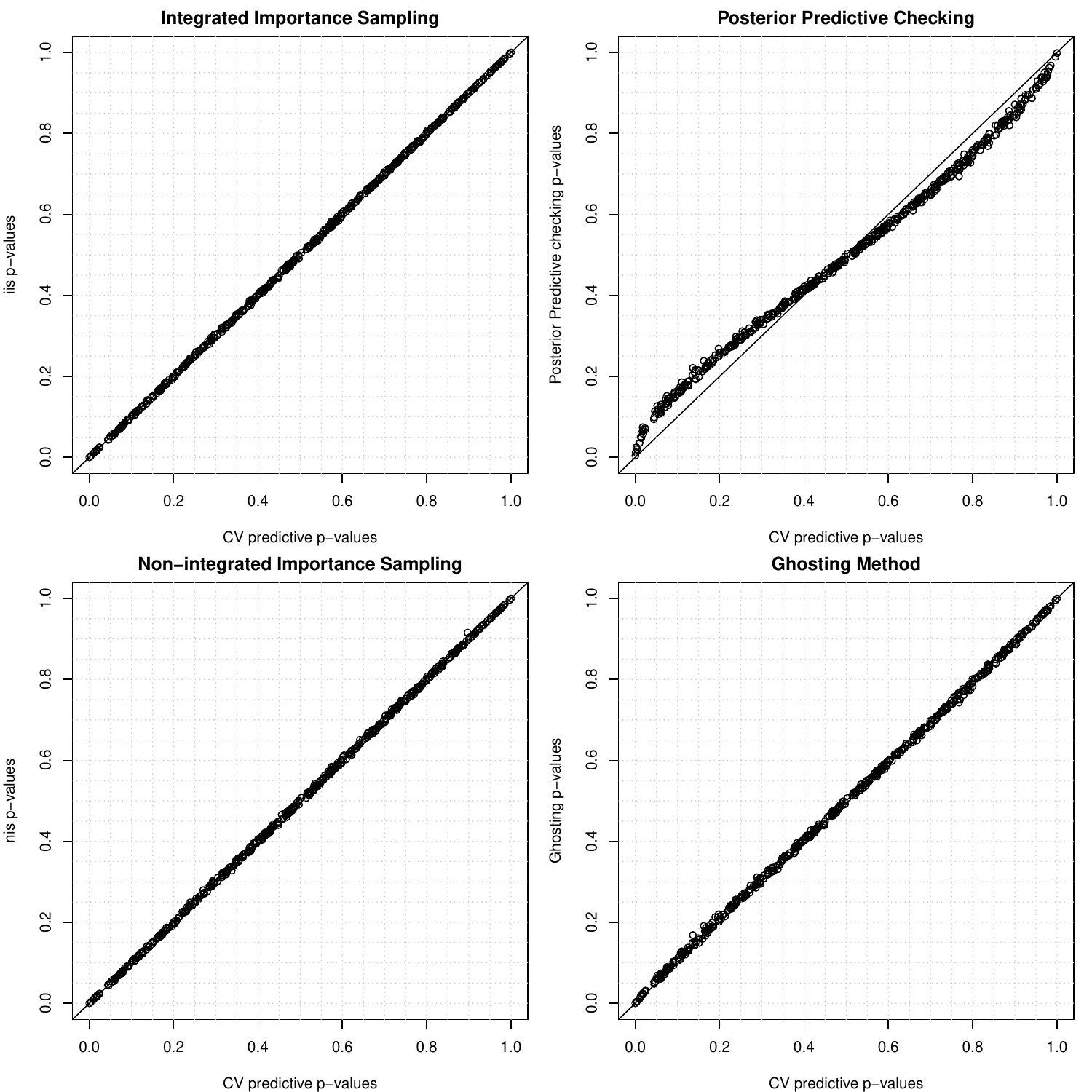}

\end{figure}

We compare the closeness of the actual LOOCV predictive p-values and the predictive p-values by the four methods using only a single MCMC fitting in Figure~\ref{fig:larynx-pvalues}. As we expect, when the sample size (number of districts) is large, the optimistic bias will decrease. From this figure, we see that all of the p-values computed with iIS, nIS and the ghosting method are very close to the actual LOOCV p-values, whereas the posterior predictive checking method, which does not consider a bias correction, still shows substantial optimistic bias (conservatism).  Although the biases of the ghosting method and the posterior predictive checking methods seem very small from Figure~\ref{fig:larynx-pvalues}, they may still lead to practical consequences. To see this, we cut each set of predictive p-values with points 0.1 and 0.9 in order to categorize the 544 districts into three pools; such categorization is needed in practice for determining which districts should be inspected further for finding out underpinning causes for the high or low residual disease rates, or for other practical decision making. Table~\ref{tab:cont-larynx} shows the two-way table of the numbers of districts in the three pools based on each set of predictive p-values computed with a single full-data MCMC fitting against the numbers obtained with the LOOCV predictive p-values.  From the table, we see that the posterior predictive checking mis-categorizes 31 and 21 districts from the category [0, 0.1) and [0.9, 1] respectively into the less extreme category [0.1, 0.9), and the ghosting method mis-categorizes 5 and 3 districts respectively.   The mis-categorization may result in omission of these districts from further inspection, which may lead to missed discovery of additional causes, e.g., certain hazard, for the disease, or leave the residents in the districts exposed to the hazard. The mis-categorization may also alter the health and research policy decisions for these districts. 

The nIS method works very well for this large dataset, making only one mis-categorization. In other words, we did not see the instability of nIS in this large dataset as in the small lip cancer dataset. This is an encouraging result for practitioners as we can see that the implementation of nIS costs neither extra time nor much extra technical effort than the posterior predictive checking. We believe that this should be generally expected when the fitted model is adequate for the data and the data size is sufficiently large, because: 1) the presence of many divergent observations causes nIS to be unstable,   and 2) omitting a single observation in a large dataset does not alter the posterior by much.   Furthermore, nIS can easily be applied to more complex models, for example models with complicated structures in both temporal and spatial domains \cite{waller1997hierarchical,lemos2009spatio-temporal} for which it is not applicable for iIS to check each observation.  In summary, nIS is a good choice when one fits a large dataset with a good model.  

One should not interpret from the good performance of nIS in large datasets that iIS is useless for large datasets. The stability of nIS depends on the relative complexity of a model to a dataset rather than the raw size of a dataset; therefore, we may not have a cheap tool to check the stability of nIS.  The stability of nIS p-values may be checked empirically by running MCMC fitting multiple times. However, from this example we see that assessing the stability of nIS by rerunning MCMC fittings costs more time (2.5hrs per MCMC running) than the time ($\approx$ 9mins) used by iIS to compute predictive p-values. In addition, there is not much more difficulty to implement iIS than the ghosting method.  Therefore,  iIS method is recommended to use in practice for both small and large datasets unless the additional computational cost becomes unaffordable in their applications. However, we notice that iIS (the ghosting method as well) is more technically complicated because of the re-generation of latent variables; this causes iIS to be more prone to implementation errors. Therefore, other predictive p-values may also be computed to check the iIS implementation. 

\begin{table}[ht]
\setlength{\tabcolsep}{3pt}
\centering
\caption{Contingency table of districts categorized by cutting predictive p-values with 0.1 and 0.9 in German larynx cancer example. The bolded numbers show the mis-categorized districts compared to CV. }\label{tab:cont-larynx}

\begin{tabular}{lrrr|rrr|rrr|rrr}
  \hline
  &\multicolumn{3}{c|}{Posterior predictive checking} & \multicolumn{3}{c|}{Ghosting method} & \multicolumn{3}{c|}{nIS}& \multicolumn{3}{c}{iIS}\\
 \multicolumn{1}{c}{CV} & [0,0.1) & [0.1,0.9) & [0.9,1] & [0,0.1) & [0.1,0.9) & [0.9,1] & [0,0.1) & [0.1,0.9) & [0.9,1] & [0,0.1) & [0.1,0.9) & [0.9,1] \\ [0pt] 
  \hline
  [0,~0.1) &  16 &  \textbf{31} &   0 &  42 &   \textbf{5} &   0 &  47 &   0 &   0 &  47 &   0 &   0 \\[0pt] 
  [0.1,~0.9) &   0 & 455 &   0 &   0 & 455 &   0 &   0 & 454 &   \textbf{1} &   0 & 455 &   0 \\[0pt] 
  [0.9,~1] &   0 &  \textbf{21} &  21 &   0 &   \textbf{3} &  39 &   0 &   0 &  42 &   0 &   0 &  42 \\[0pt]
   \hline
\end{tabular}
\end{table}

\section{Conclusions and Discussions}\label{sec:conclusions}

LOOCV predictive p-values (or a transformation) can be used for verifying the goodness-of-fit of models and for discovering systematic discrepancies between a model and a dataset. They can also be used in practice for making health and research policy decisions.  Therefore, the inaccuracy in estimating LOOCV predictive p-values may lead to wrong results in model diagnosis, and in practice may alter serious health and research policy decisions.  In this paper we have proposed to apply a new method called integrated importance sampling (iIS) for estimating LOOCV predictive p-values of disease mapping models with MCMC samples drawn from the posterior given a full dataset, as opposed to running time-consuming actual LOOCV MCMC fittings with each district removed in turn.  The innovation of our procedure lies in the strategy of integration over the random effect term (latent variable) before applying the importance sampling method to correct for the optimistic bias. These iIS predictive p-values also have the theoretical appeal of being equivalent to the LOOCV predictive p-values.  We have compared iIS with three existing methods in the literature using two real datasets. Our empirical results showed that predictive p-values estimated with iIS are in a great agreement with actual LOOCV predictive p-values in both small and large datasets. The accuracy of iIS outperformed the existing three methods---the posterior predictive checking, the ordinary importance sampling and the ghosting method when a dataset has small size relatively to the model complexity.  The iIS method requires additional (though not much) computation time and implementation effort than the posterior checking and ordinary importance sampling.  However,  the extra time and implementation effort are worthy and necessary when an application demands accurate LOOCV predictive p-values.

The iIS method can be applied to many other models with correlated or independent random effects provided that the random effect is specific to each test observation or unit. In particular, iIS can also be used in situations where a cluster of observations are collected in a unit (such as a subject with longitudinal measurements or an institution
with repeated measurements); see~\cite{marshall2007identifying} for examples. For such clustered observations, one can apply iIS to estimate a predictive p-value for each unit rather than each observation. The difficulty lies in defining a reasonable p-value for a vector of observations for measuring the tail divergence of a unit. An interesting definition of such p-value is given by~\cite{marshall2007identifying} who propose to use a latent variable (or ``parameter'') to summarize the multiple observations based on fitting a model for the clustered observations. Applications of the iIS method to the clustered observations are important and interesting in both practice and  theory. 

The applicability of iIS requires that the random effect is specific to each test observation or unit. It is not applicable when we are interested in computing a p-value for each observation in a cluster which share a latent variable. It is an interesting topic to extend the idea in iIS and the ghosting method (re-generating latent variables) to such models. However, the extension may result in a more complex formula than~\eqref{eqn:iseM}.  On the other hand, our empirical results show that the ordinary importance sampling that can be applied to a wider range of problems is a good alternative than the widely used posterior checking method. To use ordinary importance sampling, one can try a recent proposal of using a Pareto distribution to model the large importance weights~\cite{vehtari2015pareto} for reducing the instability.

\section*{Acknowledgement}

This work was supported by fundings from Natural Sciences and Engineering Research Council of Canada, and Canadian Foundation for Innovation. The authors are grateful to the editor of SIM, an associate editor, and an anonymous referee. Their comments have significantly improved the previous drafts. The authors are also grateful to Matthew Schmirler for carefully proofreading this article.


\appendix

\section{A complete tabular of the estimated predictive p-values for Scottish lip cancer data}
\setcounter{table}{0}
\renewcommand{\thetable}{A\arabic{table}}
 \begin{table}[htp]
\caption{Estimated predictive p-values$(y\obs_i)$ for the 56 districts in the Scottish lip cancer data. (Abbreviations: CV: actual leave-one-out cross-validation, PCH: posterior predictive checking, GHO: Ghosting, nIS: non-integrated (ordinary) importance sampling, and iIS: integrated importance sampling).}\label{tab:postinthree}
\centering
\begin{tabular}{|r|ccccc|r|ccccc|}\hline
ID & LOOCV & PCH & GHO & nIS & iIS & ID &LOOCV & PCH & GHO & nIS & iIS \\
\hline
 1 & 0.308 & 0.417 & 0.310 & 0.319 & 0.307 & 29 & 0.667 & 0.547 & 0.651 & 0.631 & 0.664 \\ 
  2 & 0.033 & \textbf{0.320} & \textbf{0.050} & \textbf{0.074} & 0.030 & 30 & 0.260 & 0.367 & 0.278 & 0.263 & 0.262 \\ 
  3 & 0.090 & 0.325 & 0.096 & 0.089 & 0.090 & 31 & 0.275 & 0.359 & 0.283 & 0.262 & 0.274 \\ 
  4 & 0.418 & 0.437 & 0.423 & 0.430 & 0.417 & 32 & 0.816 & 0.601 & 0.799 & 0.768 & 0.818 \\ 
  5 & 0.139 & 0.357 & 0.155 & 0.159 & 0.140 & 33 & 0.469 & 0.455 & 0.467 & 0.466 & 0.463 \\ 
  6 & 0.512 & 0.463 & 0.512 & 0.458 & 0.514 & 34 & 0.188 & 0.317 & 0.211 & 0.189 & 0.190 \\ 
  7 & 0.060 & 0.312 & 0.072 & \textbf{0.041} & 0.058 & 35 & 0.370 & 0.414 & 0.372 & 0.364 & 0.370 \\ 
  8 & 0.113 & 0.313 & 0.114 & 0.112 & 0.112 & 36 & 0.151 & 0.284 & 0.162 & 0.154 & 0.149 \\ 
  9 & 0.267 & 0.386 & 0.281 & 0.261 & 0.271 & 37 & 0.596 & 0.524 & 0.590 & 0.598 & 0.601 \\ 
  10 & 0.269 & 0.405 & 0.279 & 0.300 & 0.267 & 38 & 0.071 & 0.221 & 0.092 & 0.076 & 0.073 \\ 
  11 & 0.127 & 0.334 & 0.137 & 0.138 & 0.122 & 39 & 0.820 & 0.627 & 0.794 & 0.804 & 0.821 \\ 
  12 & 0.514 & 0.458 & 0.518 & 0.445 & 0.515 & 40 & 0.182 & 0.285 & 0.192 & 0.181 & 0.178 \\ 
  13 & 0.484 & 0.433 & 0.485 & 0.412 & 0.479 & 41 & 0.376 & 0.413 & 0.384 & 0.375 & 0.376 \\ 
  14 & 0.474 & 0.455 & 0.472 & 0.451 & 0.477 & 42 & 0.991 & \textbf{0.853} & 0.977 & 0.987 & 0.992 \\ 
  15 & 0.061 & 0.280 & 0.070 & 0.056 & 0.062 & 43 & 0.880 & 0.699 & 0.872 & 0.866 & 0.883 \\ 
  16 & 0.578 & 0.496 & 0.571 & 0.540 & 0.578 & 44 & 0.599 & 0.532 & 0.585 & 0.588 & 0.593 \\ 
  17 & 0.609 & 0.473 & 0.602 & 0.536 & 0.606 & 45 & 0.962 & \textbf{0.798} & \textbf{0.904} & 0.973 & 0.971 \\ 
  18 & 0.138 & 0.303 & 0.146 & 0.144 & 0.136 & 46 & 0.802 & 0.664 & 0.788 & 0.807 & 0.802 \\ 
  19 & 0.369 & 0.422 & 0.378 & 0.373 & 0.366 & 47 & 0.510 & 0.470 & 0.506 & 0.506 & 0.511 \\ 
  20 & 0.271 & 0.366 & 0.277 & 0.245 & 0.271 & 48 & 0.687 & 0.598 & 0.684 & 0.692 & 0.688 \\ 
  21 & 0.133 & 0.309 & 0.139 & 0.127 & 0.129 & 49 & 0.987 & \textbf{0.865} & \textbf{0.949} & 0.983 & 0.987 \\ 
  22 & 0.734 & 0.572 & 0.695 & 0.700 & 0.744 & 50 & 0.954 & \textbf{0.819} & \textbf{0.930} & 0.951 & 0.955 \\ 
  23 & 0.382 & 0.427 & 0.390 & 0.381 & 0.384 & 51 & 0.590 & 0.519 & 0.586 & 0.581 & 0.591 \\ 
  24 & 0.106 & 0.278 & 0.140 & 0.118 & 0.109 & 52 & 0.574 & 0.512 & 0.571 & 0.576 & 0.575 \\ 
  25 & 0.075 & 0.259 & 0.093 & 0.079 & 0.073 & 53 & 0.757 & 0.657 & 0.748 & 0.750 & 0.757 \\ 
  26 & 0.049 & \textbf{0.224} & \textbf{0.061} & \textbf{0.052} & 0.048 & 54 & 0.847 & 0.739 & 0.837 & 0.841 & 0.847 \\ 
  27 & 0.244 & 0.348 & 0.250 & 0.248 & 0.244 & 55 & 0.990 & \textbf{0.923} & 0.987 & 0.990 & 0.991 \\ 
  28 & 0.305 & 0.383 & 0.315 & 0.302 & 0.308 & 56 & 0.841 & 0.728 & 0.833 & 0.826 & 0.842 \\ 
  \hline
\end{tabular}
\end{table}

\section{Link to R code for MCMC fitting and computing predictive p-values}
The R code for computing predictive p-values using the four methods is available with this URL: \\
\centerline{\url{http://math.usask.ca/longhai/software/dmpvalues/dmpvalues-larynx.R}}
The instruction for downloading the dataset is included in the above file. 

\clearpage

\bibliographystyle{wileyj}
\bibliography{library}

\end{document}